# The Devil is in the Tails: Regression Discontinuity Design with Measurement Error in the Assignment Variable[*]

Zhuan Pei[1]
Cornell University and IZA

Yi Shen[2]
University of Waterloo

August 2016

**Abstract**

Identification in a regression discontinuity (RD) research design hinges on the discontinuity in the probability of treatment when a covariate (assignment variable) exceeds a known threshold. When the assignment variable is measured with error, however, the discontinuity in the relationship between the probability of treatment and the observed mismeasured assignment variable may disappear. Therefore, the presence of measurement error in the assignment variable poses a direct challenge to treatment effect identification. This paper provides sufficient conditions to identify the RD treatment effect using the mismeasured assignment variable, the treatment status and the outcome variable. We prove identification separately for discrete and continuous assignment variables and study the properties of various estimation procedures. We illustrate the proposed methods in an empirical application, where we estimate Medicaid takeup and its crowdout effect on private health insurance coverage.

[*]We thank Orley Ashenfelter, Eric Auerbach, Matias Cattaneo, Eleanor Choi, Damon Clark, Kirill Evdokimov, Hank Farber, Marjolaine Gauthier-Loiselle, Bo Honoré, Jian Kang, Marta Lachowska, David Lee, Lars Lefgren, Pauline Leung, Jia Li, Andrew Marder, Alex Mas, Jordan Matsudaira, Alexander Meister, Stephen Nei, Andrew Shephard and Lara Shore-Sheppard for illuminating discussions. We have also benefited from helpful suggestions given by the participants of the Princeton Labor Seminar, Princeton Political Science Methodology Seminar and the Advances in Econometrics conference at the University of Michigan. We thank David Card and Lara Shore-Sheppard for graciously sharing their data. Zhuan Pei gratefully acknowledges financial support from the Richard A. Lester Fellowship. Finally, we thank Yue Fang, Suejin Lee, Katherine Wen and Le Wu for excellent research assistance.

[1]Department of Policy Analysis and Management, 134 Martha Van Rensselaer Hall, Cornell University, Ithaca, NY 14853. E-mail: zhuan.pei@cornell.edu

[2]Department of Statistics and Actuarial Science, University of Waterloo, Mathematics 3 Building, 200 University Avenue West, Waterloo, ON N2L 3G1. E-mail: yi.shen@uwaterloo.ca

# 1  Introduction

Over the past two decades, studies in economics have relied on the Regression Discontinuity (RD) Design to evaluate the effects of a wide range of policy programs.[1] In an RD design, treatment is assigned based on whether an observed covariate (called the "assignment" or "running" variable) exceeds a known threshold. Provided that agents just above and just below the threshold share the same baseline characteristics and only differ in their treatment status, any difference in the outcomes between these two groups can be attributed to the causal effect of the treatment.

A classical RD design depends crucially on the econometrician's ability to observe an accurate measure of the assignment variable. In many cases, however, only a noisy version of the assignment variable is observed. This scenario is likely to occur when survey data are used, in which the value of the assignment variable comes from self-reporting as opposed to an administrative source.

A typical example is an application that uses income as an assignment variable to study the effect of means-tested transfer programs where eligibility depends on whether income falls below a certain threshold. However, most administrative data cannot be used for an RD because they only include the treatment population, namely those who enroll in the program, and contain little information on the various outcomes for applicants who are denied benefits. Therefore, practitioners may be forced to rely on survey data in order to apply an RD design. For instance, Schanzenbach (2009) uses the Early Childhood Longitudinal Study to study the effect of school lunch on obesity and compares obesity rates for children below and above the reduced-price lunch cutoff for the National School Lunch Program.[2] Hullegie and Klein (2010) study the effect of private insurance on health care utilization in the German Socio Economic Panel by using a policy rule that obliges workers with income below a threshold to participate in the public health insurance system. Koch (2013) uses the Medical Expenditure Panel Survey (MEPS) to study health insurance crowdout by focusing on income cutoffs in the Children's Health Insurance Program (CHIP). De La Mata (2012) estimates the effect of Medicaid/CHIP coverage on children's health care utilization and health outcomes with the Panel Study of Income Dynamics (PSID) and its Child Development Study (CDS) supplement.

The above studies all use income data gathered from surveys as their assignment variable in their RD analyses, but measurement error in survey data has been widely documented and studied (see Bound et al. (2001) for a review). Yet, the presence of measurement error in the assignment variable directly threatens the source of identification in an RD design, which hinges on the discontinuous relationship between the treatment and assignment variables. Even if there is perfect compliance with the discontinuous rule, there may *not* be a discontinuity in the probability of treatment conditional on the *observed* noisy assignment variable. Instead of a step function, the first-stage relationship – the probability of treatment conditional on the noisy assignment variable – is smoothly S-shaped. This lack of discontinuity may cast doubt on the identification and estimation of the program effect.

---

[1] See Lee and Lemieux (2010) for a survey of the RD literature.
[2] In addition to the RD approach, Schanzenbach (2009) also conducts analyses using other research designs.

In this paper, we study the identification of the RD treatment effect in the presence of classical measurement error in the assignment variable. We consider separately the cases of discrete and continuous assignment variables and measurement error. In the discrete case, we show that when the assignment variable is bounded (e.g., binned income), not only can we identify the first stage and outcome conditional expectation functions without specifying the measurement error distribution, but we can also identify the true assignment variable distribution by exploiting the tail behavior of the observed assignment variable distribution within the treatment and control groups. This is advantageous in the RD context, since a key appeal is the design's testability, which entails the smoothness of the true assignment variable distribution. In addition, recovery of the true assignment variable distribution may allow for the discovery of bunching and the estimation of meaningful economic quantities, even when the RD design is invalid.[3] The identification result leads to a simple minimum-distance estimation procedure for the true assignment variable distribution, and a subsequent application of Bayes' Rule allows the estimation of the first-stage and outcome relationships. Following standard reasoning, the resulting estimators are efficient, $\sqrt{N}$-consistent, and asymptotically normal.

We also explore the case where the assignment variable is continuous and propose three alternative approaches to semiparametrically or nonparametrically identify the RD treatment effect. The first approach assumes normal measurement error while remaining agnostic about the true assignment variable distribution; identification follows from the result of Schwarz and Bellegem (2010). The second approach is a novel identification-at-infinity strategy that exploits the tail behavior of the observed assignment variable distribution. The third approach adapts the nonparametric simulation-extrapoloation (SIMEX) method of Carroll et al. (1999), which assumes that the variance of the measurement error is known. The last two approaches do not identify the true assignment variable distribution, but recover the RD treatment effect parameter under perfect compliance.

---

[3]As a caveat, however, the detection of nonsmoothness or bunching in the assignment variable distribution when the variable itself is discrete will not entail a nonparametric procedure. Rather, it will depend on the parametric functional form subjectively specified by the researcher.

We illustrate our proposed methods in an empirical application where we study the takeup of Medicaid, a public health insurance program in the U.S., and the extent to which it crowds out private health insurance. Since Medicaid eligibility is only available to families with income below a strict cutoff, we apply an RD design exploiting this policy discontinuity. The data come from Card and Shore-Sheppard (2004), which are derived from the 1990-93 Survey of Income and Program Participation (SIPP). We show that the noisy income measures in SIPP indeed lead to a smooth first stage relationship between Medicaid coverage and income, and then apply two modeling approaches to recover the true income distribution, first stage, and outcome relationships. The first approach discretizes income and uses the proposed minimum distance estimation strategy, which does not impose strong functional form assumptions on the distribution of true income and the measurement error. The second approach treats income as continuous and adopts a parametric maximum likelihood estimation framework. Although the discrete approach yields noisier estimates, both methods result in similar findings: Medicaid takeup rate for the barely eligible is between 10 and 25 percent, and there is little evidence of sorting around the threshold and private insurance crowdout.

Several papers have considered the problem of measurement error in RD designs. Hullegie and Klein (2010) adopt a parametric Berkson measurement error specification, in which the true assignment variable is the sum of the observed assignment variable and an independent normally distributed measurement error. This specification is attractive in that it can be easily implemented in practice, but it implies that the distribution of the true assignment variable is smooth and precludes the testing of density discontinuity. In this paper, we adopt the more conventional classical measurement error model (Bound et al. (2001)), which allows nonsmoothness in the assignment variable density. Dong (2015) considers rounding (nonclassical) error in the assignment variable typically encountered in age-based RD designs. In three recent studies, Yu (2012) and Yanagi (2014) consider the identification problem assuming "small" measurement error variance; Davezies and Le Barbanchon (2014) assume the availability of an auxiliary dataset so that the true assignment variable distribution can be observed (for the treated population). In contrast to these studies, the measurement error distribution in our set up is not restricted to have small second moments, nor do we assume the researchers can access an auxiliary dataset.[4]

The remainder of the paper is organized as follows. Section 2 introduces the statistical model. Section

---

[4] A related but distinct problem is heaping in the assignment variable (Almond et al. (2010), Almond et al. (2011), Barreca et al. (2011) and Barreca et al. (2016)). In the heaping setup, treatment assignment is based on the observed value of the assignment variable. The problem at hand is one where we do not observe the variable determining treatment.

3 discusses the case where the assignment variable and measurement error are discrete, and establish identification and estimation procedures of the true assignment variable distribution as well as that of the RD treatment effect parameter under both the sharp and fuzzy design. Section 4 studies the case of continuous assignment variable and measurement error, proposes three identification approaches, and investigates the behavior of various estimation strategies. Section 5 provides an empirical illustration by employing the proposed methods to estimate Medicaid takeup and crowdout. Section 6 concludes.

## 2 Baseline Statistical Model

In a conventional sharp RD design, the econometrician observes assignment variable $X^*$, eligibility/treatment $D^*$ and outcome $Y$ where

$$Y = y(D^*, X^*, \varepsilon)$$
$$D^* = 1_{[X^*<0]} \quad (1)$$

$y$ is a function continuous in its second argument, $\varepsilon$ is the error term unobserved to the econometrician, and the eligibility threshold is normalized to zero.[5] A standard result (e.g. Hahn et al. (2001)) is that the treatment effect $\delta_{sharp} = E[y(1,0,\varepsilon) - y(0,0,\varepsilon)|X^* = 0]$ is identified as

$$\delta_{sharp} = \lim_{x^*\uparrow 0} E[Y|X^* = x^*] - \lim_{x^*\downarrow 0} E[Y|X^* = x^*]$$

when conditional expectation of the response function $E[y(D^*,X^*,\varepsilon)|X^* = x^*]$ is continuous at $x^* = 0$ for $D^* = 0,1$.

In this paper, we consider the extension where $X^*$ is not observed but a noisy measure of it is. Let $X$ be the observed assignment variable, and $u \equiv X - X^*$ the measurement error. As mentioned in section 1, a key assumption that distinguishes this study from Hullegie and Klein (2010) is that the measurement error is independent of the true assignment variable as opposed to the observed assignment variable. Formally,

**Assumption 1 (Independence).** $X^* \perp\!\!\!\perp u$

---

[5]Note that in some applications $D^* = 1_{[X^* \geqslant 0]}$ is the treatment determining mechanism. However, most of the motivating examples in section 1 follow $D^* = 1_{[X^*<0]}$, i.e. program eligibility depends on whether the assignment variable (income) falls *below* a known cutoff.

As is well known, classical measurement error in an ordinary least squares (OLS) regressor leads to attenuation bias. So a natural question is whether we can characterize the estimated RD treatment in a similar way. The answer is both yes and no. On the one hand, continuous measurement error completely smooths out the first stage (and outcome) discontinuity as seen in Figure 1. It follows that the attenuation factor is *zero* for both the first-stage and outcome, and it can be regarded as the most severe form of attenuation. On the other hand, it is difficult to discuss the bias of the fuzzy RD estimand, which is simply undefined due to the lack of first stage discontinuity. However, it is possible to naively perform local regressions on both sides of the threshold and obtain nonzero discontinuity estimates for both the first stage and the outcome through misspecification. Whether the resulting ratio under- or over-states the true RD treatment effect is anyone's guess.

The issue of attenuation, or the lack thereof, can also be seen by following an alternative definition of the sharp RD estimand: $\tilde{\delta}_{sharp} = \lim_{x\uparrow 0}[Y|X=x, D^*=1] - \lim_{x\downarrow 0}[Y|X=x, D^*=0]$. When $X$ is measured without error, $\tilde{\delta}_{sharp}$ coincides with the conventional sharp RD estimand, $\lim_{x\uparrow 0} E[Y|X=x] - \lim_{x\downarrow 0} E[Y|X=x]$. However, unlike the conventional RD estimand, which attenuates to zero even when $X$ is measured with very small error, $\tilde{\delta}_{sharp}$ gradually drifts apart from $\delta_{sharp}$ as the measurement error variance increases. In fact, this is an attractive property of $\tilde{\delta}_{sharp}$, which we exploit in section 4. For now, we argue that $\tilde{\delta}_{sharp}$ is not necessarily attenuated, and the direction of the bias in $\tilde{\delta}_{sharp}$ depends on the particular functional form in the relationship between $X^*$ and $Y$. Suppose, for example, $Y$ is linear in $X^*$ for both the treatment and control groups: $E[Y|X^*, D^*=0] = \beta_0 + \beta_1 X^*$ and $E[Y|X^*, D^*=1] = \beta_2 + \beta_3 X^*$. If we project $Y$ on $X$ within the treatment and control groups, standard OLS results indicate that the slope estimators $\hat{\beta}_1$ and $\hat{\beta}_3$ are both attenuated. This in turn implies that if $\beta_1, \beta_3 < 0$, then $\hat{\beta}_2$ is biased upward and $\hat{\beta}_0$ downward, and the resulting RD estimator $\hat{\beta}_2 - \hat{\beta}_0$ is biased upward. When $\delta_{sharp} = \beta_2 - \beta_0 > 0$, we actually arrive at an exaggerated, rather than attenuated, RD estimate! In short, the association between measurement error and attenuation comes from slope estimation in an error-in-variable and does not generalize to the intercept.

Given the challenge of not directly observing $X^*$, this paper studies whether we can recover the RD treatment effect with classical measurement error. The first step in our quest is the identification of the distribution of $X^*$. As mentioned in the introduction, not only is the identification of the $X^*$ distribution used to recover of the RD treatment effect parameter $\delta_{sharp}$, but it can also be used for testing the validity of the RD design. However, it is not possible to identify the distribution of $X^*$ from the observed distribution of $X$ absent any other information. In the presence of measurement error, economists have traditionally proposed using repeated and possibly noisy measures of the true explanatory variable (e.g. Ashenfelter and Krueger (1994), Black et al. (2000), Hausman et al. (1991), Li and Vuong (1998), Schennach (2004)). However, such measures may not be available in the data. What is helpful in the RD context is that observed program eligibility $D^* = 1_{[X^*<0]}$, which is a deterministic function of $X^*$, is informative of the value of $X^*$. Therefore, it becomes an interesting question as to whether and under what additional assumptions is the distribution of $X^*$ identifiable from the joint distribution of $(X, D^*)$.[6]

This question is addressed in the subsequent sections. We discuss in section 3 the identification of the assignment variable distribution and the RD treatment effect parameter under the assumption that $X^*$ and $u$ are discrete. Sufficient conditions for identification are provided for the case in which $X^*$ and $u$ have bounded support, and an example is constructed in section C of the Appendix to show that the model is, in general, not identified when the bounded support assumption of $X^*$ and $u$ is relaxed.

A discrete assignment variable setup may appear odd given the continuity assumption of $E[y(D^*, X^*, \varepsilon)|X^* = x^*]$ in identifying the treatment effect in a sharp RD design, but it is necessary in many policy contexts where an RD design appears compelling (Lee and Card (2008)). Even if the assignment variable is continuous (e.g. income), the discretization of $X^*$ can be thought of as a binned-up approximation (this is common practice in graphical presentations of most RD applications). We nevertheless point out that if we assume independence between the underlying *continuous* assignment variable and measurement error, then their respective discretized versions are not going to be independent.[7] However, we can also start with the assumption that the discretized $X^*$ and $u$ are independent, in which case their continuous versions will not be. Which one of these assumptions is correct? We are inclined to believe that *neither* of these assumptions is likely to hold empirically, but it is important to gauge whether they reasonably approximate the data. We discuss an overidentification test to that effect in section 3 and an empirical illustration in section 5. It is also worth noting that when the discrete $X$ and $X^*$ have the same support (e.g. whey then denote age/birthdate, student enrollment), the measurement error is not independent to the true assignment variable – the extreme form of this is the misclassification error in a binary variable, which is known to be mean-reverting. Identification

---

[6]As pointed out by a referee, it is possible that the RD threshold is also measured with error. For example, the econometrician may not have all the information to precisely construct a family's eligibility threshold for a means-tested program. While it is difficult to disentangle the measurement error in threshold from that in income or to identify the true threshold, we can still identify the RD treatment effect without being able to do so. To see this, suppose the true eligibility assignment mechanism is $D^* = 1_{[W^*<c^*]}$, where $W^*$ is, say, the actual family income and $c^*$ the monetary eligibility threshold (note that we use the normalization $X^* = W^* - c^*$ above, and the threshold is normalized to zero). Suppose the econometrician only sees proxies $W = W^* + u$ and $c = c^* + v$, but not $W^*$ and $c^*$ directly. In this case, we can rewrite the eligibility assignment mechanism as $D^* = 1_{[X-u+v<0]} = 1_{[X-\tilde{u}<0]}$. Assuming $X^* \perp\!\!\!\perp v$, it follows that $X^* \perp\!\!\!\perp \tilde{u}$, and we have a model isomorphic to (1).

[7]Dong (2015) makes a similar point in the case of age rounding.

of the assignment variable distribution as well as the RD conditional expectation function with this type of nonclassical measurement error is a corollary of the problem nicely solved by Chen et al. (2009)[8]. Under appropriate rank conditions and shape restrictions on the conditional expectation function, Chen et al. (2009) employ the observed conditional characteristic function of $Y$ and its derivative for identification.

The observability of program *eligibility*, or equivalently perfect compliance, is assumed in the sharp RD model (1). In most applied contexts (such as those in the studies cited above), however, this assumption is rarely satisfied. In all social programs, for example, the take-up rate of entitlement programs among eligible individuals and families is not 100 percent.[9] For these programs, only take-up $D$ is observed in the data, which is no longer a deterministic function of true assignment variable $X^*$. As a consequence, additional assumptions on the measurement error distribution are needed for the identification of the distribution of $X^*$, and we explore these assumptions in section 3.2. In section 3.3, we discuss how the treatment effect from the RD design can be identified by assuming nondifferential measurement error.

Finally, in section 4, we discuss identification and estimation when $X^*$ and $u$ are continuously distributed.[10] Building on the insight from the discrete case, when $u$ is assumed to be normally distributed, we show that the $X^*$ distribution and the RD treatment effect are semiparametrically identified under both perfect and imperfect compliance. When we relax normality, we can still recover the RD treatment effect under perfect compliance via two approaches: a) an identification-at-infinity strategy and b) an application of the SIMEX method of Carroll et al. (1999) when the measurement error variance is known.

## 3 Discrete Assignment Variable and Measurement Error

### 3.1 Identification of the True Assignment Variable Distribution: Perfect Compliance

In this section, we provide sufficient conditions for identifying the distributions of $X^*$ and $u$ from the joint distribution of $X$ and $D^*$ in model (1) where $X^*$ and $u$ are *discrete and bounded.* Formally, what we observe

---

[8] We thank a referee for suggesting this reference.
[9] See Currie (2006) for a survey on benefit take-up in social programs.
[10] Arguably another possibility not considered here is when $X^*$ is continuous but $u$ is discrete. A starting point in this setting is Dong and Lewbel (2011).

are

$$D^* = 1_{[X^*<0]}$$
$$X = X^* + u \qquad (2)$$

The identification result is shown in two steps: 1) the identification of the support of $X^*$ and $u$ and 2) the identification of the probability mass at each point in the support of $X^*$ and $u$. In additional to independence between $X^*$ and $u$, the identification result relies on the assumption of positive mass around the threshold 0 in the $X^*$ distribution and a technical rank condition to be discussed in detail later.

Denote the support of any random variable $Z$ by $\text{support}_Z$, and let $\min\{\text{support}_Z\} = L_Z$ and $\max\{\text{support}_Z\} = U_Z$ for a discrete and bounded $Z$. Without loss of generality, we consider the case where $\text{support}_{X^*}, \text{support}_u \subseteq \mathbb{Z}$, the set of integers. Formally, the discrete and bounded support assumption is written as

**Assumption DB (Discrete and Bounded Support).** $\text{support}_{X^*} \subseteq \{L_{X^*}, L_{X^*}+1, ..., U_{X^*}-1, U_{X^*}\}$ and $\text{support}_u \subseteq \{L_u, L_u+1, ..., U_u-1, U_u\}$ where $L_{X^*}, U_{X^*} L_u, U_u \in \mathbb{Z}$.

Abstracting from sampling error, the joint distribution of $(X, D^*)$ observed by the econometrician is fully characterized by the distribution of $X$ conditional on $D^*$ and the marginal distribution of $D^*$. The assumption of independence between $X^*$ and $u$ gives strong implications relating their respective supports to the observed support of $X$, conditional on $D^*$. Specifically,

$$\min\{\text{support}_{X|D^*=d}\} = \min\{\text{support}_{X^*|D^*=d}\} + \min\{\text{support}_u\}$$
$$\max\{\text{support}_{X|D^*=d}\} = \max\{\text{support}_{X^*|D^*=d}\} + \max\{\text{support}_u\} \text{ for } d=0,1 \qquad (3)$$

which impose four restrictions (four equations in the equation array (3)) on six unknowns ($\min\{\text{support}_{X^*|D^*=0}\}$, $\max\{\text{support}_{X^*|D^*=0}\}$, $\min\{\text{support}_{X^*|D^*=1}\}$, $\max\{\text{support}_{X^*|D^*=1}\}$, $\min\{\text{support}_u\}$ and $\max\{\text{support}_u\}$). In order to identify the supports of $X^*$ and $u$, we impose the additional assumption

**Assumption 2 (Threshold Support).** $-1, 0 \in \text{support}_{X^*}$

Assumption 2 states that there exist agents with $X^*$ right at and below the eligibility threshold 0. This is not a strong assumption and must be satisfied in all valid RD designs because the quasi-experimental variation of an RD design comes from agents around the threshold. It is straightforward to show that the addition of this weak assumption is sufficient for identifying the supports.

**Lemma 1** (**Support Identification in a Sharp Design**) Under Assumptions DB, 1 and 2, $L_{X^*}, U_{X^*}, L_u$ and $U_u$ are identified.

*Proof.* The relationship $D^* = 1_{[X^* < 0]}$ implies

1) $\min\{\text{support}_{X^*|D^*=1}\} = \min\{\text{support}_{X^*}\} = L_{X^*}$,
2) $\max\{\text{support}_{X^*|D^*=0}\} = \max\{\text{support}_{X^*}\} = U_{X^*}$
3) $\max\{\text{support}_{X^*|D^*=1}\} < 0$
4) $\min\{\text{support}_{X^*|D^*=0}\} \geqslant 0$.

By Assumption 2, $\Pr(X^* = -1|D^* = 1) = \frac{\Pr(X^*=-1)}{\Pr(D^*=1)} > 0$ and $\Pr(X^* = 0|D^* = 0) = \frac{\Pr(X^*=0)}{\Pr(D^*=0)} > 0$. Consequently, Assumption 2 translates into statements about the support of $X^*|D^* = 0$ and $X^*|D^* = 1$:

$$\max\{\text{support}_{X^*|D^*=1}\} = -1$$
$$\min\{\text{support}_{X^*|D^*=0}\} = 0$$

i.e. Assumption 2 allows us to pin down two of the six unknowns in (3). It follows that the remaining four unknowns in (3), $L_{X^*}, U_{X^*}, L_u$ and $U_u$ are now exactly identified:

$$L_u = L_{X|D^*=0}$$
$$U_u = U_{X|D^*=1} + 1$$
$$L_{X^*} = L_{X|D^*=1} - L_{X|D^*=0}$$
$$U_{X^*} = U_{X|D^*=0} - U_{X|D^*=1} - 1$$

□

Intuitively, individuals who are in the program ($D^* = 1$) but appear ineligible ($X \geqslant 0$) have a positive measurement error $u > 0$. Analogously, those with $D^* = 0$ but $X < 0$ have a negative measurement error $u < 0$. This is essentially the insight behind Lemma 1.

With the support of $X^*$ identified, we next derive the identification of the probability mass of $X^*$ at every point in its support. Denote the probability mass of $X^*$ by $p_k$ at each integer $k$, and denote that of $u$ by $m_l$.

Let the conditional probability masses of the observed assignment variable $X$ be $q_j^1 \equiv \Pr(X = j | D^* = 1)$ for $j \in \{L_{X|D^*=1}, L_{X|D^*=1}+1, ..., U_{X|D^*=1}-1, U_{X|D^*=1}\}$, $q_j^0 \equiv \Pr(X = j | D^* = 0)$ for $j \in \{L_{X|D=0}, L_{X|D=0}+1, ..., U_{X|D=0}-1, U_{X|D=0}\}$, and the marginal probabilities $r^1 \equiv \Pr(D^* = 1)$ and $r^0 \equiv \Pr(D^* = 0)$.

Under the independence assumption of $X^*$ and $u$, the distribution of $X|D^*$ is the convolution of the distribution of $X^*|D^*$ and that of $u$. In particular,

$$q_j^1 = \frac{\sum_{k<0} p_k m_{j-k}}{\sum_{k<0} p_k}$$
$$q_j^0 = \frac{\sum_{k \geq 0} p_k m_{j-k}}{\sum_{k \geq 0} p_k} \qquad (4)$$

Additionally, the marginal probabilities of $D$ give rise to two more restrictions on the parameters of interest, namely

$$r^1 = \sum_{k<0} p_k$$
$$r^0 = \sum_{k \geq 0} p_k \qquad (5)$$

Note that $r^1, r^0 > 0$ under Assumption 2, and the $q_j^1$ and $q_j^0$'s are thus well-defined. Note also that $\sum_k p_k = 1$ follows from $r^1 + r^0 = 1$ and (5), and $\sum_l m_l = 1$ follows from $\sum_j (q_j^1 r^1 + q_j^0 r^0) = 1$, and they are therefore redundant constraints.

Together, (4) and (5) represent $2K_u + K_{X^*}$ restrictions on $K_u + K_{X^*}$ parameters, where $K_{X^*} = |\{L_{X^*}, L_{X^*}+1, ..., U_{X^*}-1, U_{X^*}\}|$ and $K_u = |\{L_u, L_u+1, ..., U_u-1, U_u\}|$ denote the number of probability mass points to be identified in the $X^*$ and $u$ distributions. Even though there are more constraints than parameters, it is not clear that the $X^*$ distribution is always identified because of the nonlinearity in (4). To formally investigate the identifiability of the parameters, we introduce the following notations: Let $p_k^1 = \frac{p_k}{r^1}$ for $k < 0$ and $p_k^0 = \frac{p_k}{r^0}$ for $k \geq 0$. Define $Q^1(t) = \sum_j q_j^1 e^{tj}$, $Q^0(t) = \sum_j q_j^0 e^{tj}$, $P^1(t) = \sum_k p_k^1 e^{tk}$, $P^0(t) = \sum_k p_k^0 e^{tk}$ and $M(t) = \sum_l m_l e^{tl}$, which are the moment generating functions (MGF's) of the random variables, $X|D = 1$, $X|D = 0$, $X^*|D = 1$, $X^*|D = 0$ and $u$.[11] It is a well-known result that the moment generating function of the sum of two independent random variables is the product of the moment generating functions of the two variables (see for example Chapter 10 of Grinstead and Snell (1997)). Consequently, equations (4) and (5)

---

[11] Because of the bounded support assumption, the defined moment generating functions always exist and are positive for all $t$.

can be compactly represented as

$$
\begin{aligned}
Q^1(t) &= P^1(t)M(t) \text{ for all } t \neq 0 \\
Q^0(t) &= P^0(t)M(t) \text{ for all } t \neq 0 \\
P^1(0) &= 1 \\
P^0(0) &= 1
\end{aligned}
\qquad (6)
$$

For the first two equations above, the coefficients on the $e^{tj}$ term in $Q^1(t)$ and $Q^0(t)$ are $q_j^1$ and $q_j^0$ respectively for each $j$, and those on the $e^{tj}$ term in $P^1(t)M(t)$ and $P^0(t)M(t)$ are $\sum_k p_k^1 m_{j-k}$ and $\sum_k p_k^0 m_{j-k}$ respectively. The last two equations are simply another way of writing (5).

Because $P^1(t)$ and $P^0(t)$ are everywhere positive, (6) implies that

$$M(t) = \frac{Q^1(t)}{P^1(t)} = \frac{Q^0(t)}{P^0(t)}$$

and it follows that,

$$P^0(t)Q^1(t) = P^1(t)Q^0(t)$$

which eliminates the nuisance parameters associated the measurement error distribution.[12] Matching the coefficients in $P^0(t)Q^1(t)$ to those in $P^1(t)Q^0(t)$ along with the constraint $P^1(0) = P^0(0) = 1$ results in the

---

[12]Note that the independence assumption implies that the measurement error distribution is invariant with respect to treatment status, and it plays a key role in the derivation above. This is certainly a strong restriction, since it is possible that being in the treatment may affect measurement error, but we cannot achieve identification without the restriction.

following *linear* system of equations in terms of the $p_k^1$'s and $p_k^0$'s:

$$\underbrace{\begin{bmatrix} q_{U_u-1}^1 & 0 & \cdots & 0 & -q_{U_{X^*}+U_u}^0 & 0 & \cdots & 0 \\ q_{U_u-2}^1 & q_{U_u-1}^1 & \cdots & 0 & -q_{U_{X^*}+U_u-1}^0 & -q_{U_{X^*}+U_u}^0 & \cdots & 0 \\ \vdots & q_{U_u-2}^1 & \cdots & \vdots & \vdots & -q_{U_{X^*}+U_u-1}^0 & \cdots & \vdots \\ q_{L_u+L_{X^*}}^1 & \vdots & \cdots & 0 & -q_{L_u}^0 & \vdots & \cdots & 0 \\ 0 & q_{L_u+L_{X^*}}^1 & \cdots & q_{U_u-1}^1 & 0 & -q_{L_u}^0 & \cdots & -q_{U_{X^*}+U_u}^0 \\ \vdots & 0 & \cdots & q_{U_u-2}^1 & \vdots & 0 & \cdots & -q_{U_{X^*}+U_u-1}^0 \\ \vdots & \vdots & \cdots & \vdots & \vdots & \vdots & \cdots & \vdots \\ 0 & 0 & \cdots & q_{L_u+L_{X^*}}^1 & 0 & 0 & \cdots & -q_{L_u}^0 \\ 1 & 1 & \cdots & 1 & 0 & 0 & \cdots & 0 \\ 0 & 0 & \cdots & 0 & 1 & 1 & \cdots & 1 \end{bmatrix}}_{\mathbf{Q}:\,(K_{X^*}+K_u)\times K_{X^*}} \underbrace{\begin{bmatrix} p_{U_{X^*}}^0 \\ p_{U_{X^*}-1}^0 \\ \vdots \\ p_0^0 \\ p_{-1}^1 \\ p_{-2}^1 \\ \vdots \\ p_{L_{X^*}}^1 \end{bmatrix}}_{\mathbf{p}:\,K_{X^*}\times 1} = \underbrace{\begin{bmatrix} 0 \\ 0 \\ \vdots \\ 0 \\ 0 \\ \vdots \\ 1 \\ 1 \end{bmatrix}}_{\mathbf{b}:\,(K_{X^*}+K_u)\times 1}$$

(7)

Standard results in linear algebra can be invoked to provide identification of the probability masses. Denote system (7) with the compact notation $\mathbf{Qp} = \mathbf{b}$, where $\mathbf{Q}$ is the $(K_{X^*} + K_u) \times K_{X^*}$ data matrix, $\mathbf{p}$ is the $K_{X^*} \times 1$ parameter vector and $\mathbf{b}$ is the $(K_{X^*} + K_u) \times 1$ vector of 0's and 1's. The parameter vector $\mathbf{p}$ is identified if and only if $\mathbf{Q}$ is of full rank. Note that there are more rows than columns in $\mathbf{Q}$, i.e. $K_{X^*} + K_u > K_{X^*}$, and therefore we introduce the following assumption

**Assumption 3 (Full Rank).** $\mathbf{Q}$ in equation (7) has full column rank.

Note that Assumption 3 is not always satisfied, and an example is provided in section A of the Appendix. At the same time, Assumption 3 is directly testable because $\text{rank}(\mathbf{Q}) = \text{rank}(\mathbf{Q}^T\mathbf{Q})$, and $\mathbf{Q}$ is of full rank if and only if $\det(\mathbf{Q}^T\mathbf{Q}) \neq 0$. The distribution of the determinant estimator can be obtained by a simple delta method because it is a polynomial in the $q_j^0$'s and $q_j^1$'s, the observed probability masses.

With Assumption 3, the $\mathbf{p}$ vector is identified. Because $p_k = r^1 p_k^1$ for $k < 0$ and $p_k = p_k^0 r^0$ for $k \geqslant 0$ and because $r^1$ and $r^0$ are observed, uniqueness of the $p_k^1$'s and the $p_k^0$'s implies the uniqueness of the $p_k$'s. Although parameters of the the measurement error distribution are eliminated in (7), they are identified after the identification of the $p_k$'s as shown in section B of the Appendix. Formally, the identification of probability masses is summarized in the following Lemma.

**Lemma 2 (Probability Mass Identification in a Sharp Design)** Suppose $(\{p_k\}, \{m_l\})$ $(k \in \{L_{X^*}, L_{X^*} + 1, ..., U_{X^*} - 1, U_{X^*}\}, l \in \{L_u, L_u + 1, ..., U_u - 1, U_u\})$ solves the system of equations consisting of (4) and (5). Then $(\{p_k\}, \{m_l\})$ is the unique solution if and only if Assumption 3 holds.

Combining Lemma 1 and 2 implies the identification of model (2):

**Proposition 1** Under Assumption DB, 1, 2, and 3, the distributions of $X^*$ and $u$ are identified.

Intuitively, the identification result extracts information about $u$ from the tails of the $X^*$ distribution in the treatment and control groups. The discrete and bounded assumption reduces the dimensionality of the identification problem and fully specifies the statistical model with a finite number of parameters, even though we are completely agnostic about the shapes of the $X^*$ and $u$ distributions over their respective support. After we have pinned down the set of parameters, we identify the $X^*$ distribution by starting from the tails and working our way in, using the independent measurement error assumption.

When we relax boundedness, we no longer have the luxury of dealing with only a finite number of parameters. Consequently, model (2) is not identified, in general, as shown through a constructive example in section C of the Appendix.

## 3.2 Identification of the True Assignment Variable Distribution: Imperfect Compliance

As mentioned in section 2, the assumption of perfect compliance or equivalently the observability of eligibility ($D^* = 1_{[X^* < 0]}$) is not often satisfied, as is the case in almost all means-tested social programs. Instead, only a measure of program participation $D$ may be available. In this section, we consider the more realistic case of imperfect compliance for discrete and bounded assignment variable $X^*$ and measurement error $u$. The task becomes the identification of the $X^*$ distribution from the observed joint distribution $(X, D)$.

Rather than having benefit receipt $D$ as a deterministic step function of $X^*$, $\Pr(D = 1 | X^*)$ is potentially an unrestricted function in $X^*$ even though eligibility is still given by $D^* = 1_{[X^* < 0]}$. In the extreme, program participation $D$ could be independent from $X^*$ and therefore would not provide any information for $X^*$. If this were the case, deconvolving $X^*$ and $u$ from the observed joint distribution $(X, D)$ would not be possible. In many programs (typically means-tested transfer programs), however, it is the case that if an agent's true assignment variable is above the eligibility threshold, she is forbidden from participating in the program, that is

**Assumption 4 (One-sided Fuzzy).** $\Pr(D=1|X^*=x^*)=0$ for $x^* \geqslant 0$.

Assumption 4 informs the researcher that any agent with $D=1$ has true assignment variable $X^* < 0$. It follows that the upper end point in the $X|D=1$ distribution identifies $U_u$ (the upper end point of the $u$ distribution as defined in 3.1) provided that Assumptions 1 and 2 hold for the $D=1$ population. Unlike in the perfect compliance scenario, where $X^* \perp\!\!\!\perp u$ conditional on $D^*$ directly follows from Assumption 1, an additional assumption is needed to ensure the independence between $X$ and $u$ conditional on $D=1$:

**Assumption 1F (Strong Independence).** $u \perp\!\!\!\perp X^*, D$.

and the required extension of Assumption 2 is

**Assumption 2F (Threshold Support: Fuzzy).** $-1 \in \text{support}_{X^*|D=1}$ and $0 \in \text{support}_{X^*}$.

We make two remarks regarding Assumptions 1F and 2F. First, note that a weaker version of Assumption 1F, that $u \perp\!\!\!\perp X^*$ conditional on $D$, suffices for the results below. However, it may be difficult to find situations where this weaker assumption is empirically justified but Assumption 1F is not. In fact, this assumption does not even imply that $X^* \perp\!\!\!\perp u$ unconditionally, a standard assumption in the measurement error literature. Therefore, we propose the stronger Assumption 1F. Second, even though Assumption 2F is stronger than Assumption 2, it needs to be satisfied in a valid fuzzy RD design without measurement error. There must be nonzero take-up just below the eligibility cutoff, without which a first-stage discontinuity does not exist.

Even with Assumptions IF and 2F, one still needs to distinguish between nontakeup and ineligibility. That is, an agent with $D=0$ and $X=-1$ could have true income $X^*=1$ (with an implied measurement error $u=-2$) and is not program eligible; or she could be eligible with income $X^*=-1$ (with an implied measurement error $u=0$) but chooses not to participate in the program. On the one hand, if every observation with $D=0$ is treated as ineligible, then the lower end point in the support of $u$, $L_u$, is that in the $X|D=0$ distribution. On the other hand, if every observation with $D=0$ is treated as an eligible nontakeup, then $L_u$ is 0. Clearly, the two treatments imply different distributions. However, if the researcher believes that the identified length of the right tail in the $u$ distribution sheds light on the length of its left tail, it may be reasonable to assume

**Assumption 5 (Symmetry in Support).** $L_u = -U_u$

which is weaker than imposing symmetry in the measurement error distribution as is conventional in the literature.

With the additional Assumptions 4, 1F, 2F and 5, the supports of the $X^*$ and $u$ are identified:

**Lemma 1F** (**Support Identification in a Fuzzy Design**) Under Assumptions DB, 1F, 2F, 4 and 5, the upper and lower end points of the $u$ distribution are given by

$$
\begin{aligned}
U_u &= U_{X|D=1} + 1 \\
L_u &= -(U_{X|D=1} + 1)
\end{aligned}
\tag{8}
$$

and those of the $X^*|D = d$ ($d = 0, 1$) distributions are given by:

$$
\begin{aligned}
L_{X^*|D=1} &= L_{X|D=1} - L_u \\
L_{X^*|D=0} &= L_{X|D=0} - L_u \\
U_{X^*|D=1} &= -1 \\
U_{X^*|D=0} &= U_{X|D=0} - U_u
\end{aligned}
\tag{9}
$$

As in section 3.1, the right tail of $X^* \geqslant 0$ and $D = 1$ population provides information on the length of the right tail of the measurement error distribution thanks to Assumption 4. The length of the left tail of the measurement error distribution is then obtained following Assumption 5. As it turns out, the identification of probability masses can proceed analogously as in section 3.1.

Because of the existence of nonparticipants, however, the distribution of $X^*$ conditional $D = 0$ is also supported on negative integers. The number of parameters therefore is larger than that in the perfect compliance case even if the support of the unconditional $X^*$ distribution does not change. It is straightforward to show that the convolution relationships under Assumption 1F again lead to a system of equations

$\mathbf{Q_F p_F = b_F}$:

$$
\begin{bmatrix}
q^1_{U_u-1} & 0 & \cdots & 0 & -q^0_{U^0_{X^*}+U_u} & 0 & \cdots & 0 \\
q^1_{U_u-2} & q^1_{U_u-1} & \cdots & 0 & -q^0_{U^0_{X^*}+U_u-1} & -q^0_{U_{X^*}+U_u} & \cdots & 0 \\
\vdots & q^1_{U_u-2} & \cdots & \vdots & \vdots & -q^0_{U_{X^*}+U_u-1} & \cdots & \vdots \\
q^1_{L_u+L^1_{X^*}} & \vdots & \cdots & 0 & -q^0_{L^0_{X^*}+L_u} & \vdots & \cdots & 0 \\
0 & q^1_{L_u+L^1_{X^*}} & \cdots & q^1_{U_u-1} & 0 & -q^0_{L^0_{X^*}+L_u} & \cdots & -q^0_{U_{X^*}+U_u} \\
\vdots & 0 & \cdots & q^1_{U_u-2} & \vdots & 0 & \cdots & -q^0_{U_{X^*}+U_u-1} \\
\vdots & \vdots & \cdots & \vdots & \vdots & \vdots & \cdots & \vdots \\
0 & 0 & \cdots & q^1_{L_u+L^1_{X^*}} & 0 & 0 & \cdots & -q^0_{L^0_{X^*}+L_u} \\
1 & 1 & \cdots & 1 & 0 & 0 & \cdots & 0 \\
0 & 0 & \cdots & 0 & 1 & 1 & \cdots & 1
\end{bmatrix}
\begin{bmatrix}
p^0_{U^0_{X^*}} \\
p^0_{U^0_{X^*}-1} \\
\vdots \\
p^0_{L^0_{X^*}+1} \\
p^0_{L^0_{X^*}} \\
p^1_{-1} \\
p^1_{-2} \\
\vdots \\
p^1_{L^1_{X^*}}
\end{bmatrix}
=
\begin{bmatrix}
0 \\
0 \\
\vdots \\
0 \\
0 \\
1 \\
1
\end{bmatrix}
$$

$\underbrace{\phantom{XXX}}_{\mathbf{Q_F}: (K^F_{X^*}+K_u) \times K^F_{X^*}}$ $\underbrace{\phantom{X}}_{\mathbf{p_F}: K^F_{X^*} \times 1}$ $\underbrace{\phantom{X}}_{\mathbf{b_F}: (K^F_{X^*}+K_u) \times 1}$

(10)

Note that in (10), we

1) adopt the notation $L_{X^*|D=d} = L^d_{X^*}$, $U_{X^*|D=d} = U^d_{X^*}$ for $d = 0, 1$;

2) define $K^F_{X^*} = U^0_{X^*} - L^0_{X^*} - L^1_{X^*} + 1$ and

3) use the superscript 1 and 0 to indicate conditioning on $D = 1$ and $D = 0$ (as opposed to $D^* = 1$ and $D^* = 0$ in (7)).

Analogous to (7), the number of rows in $\mathbf{Q_F}$ is $K_u$ more than the number of columns. Full column rank in $\mathbf{Q_F}$ is again a necessary and sufficient condition for identification:

**Assumption 3F (Full Rank: Fuzzy).** $\mathbf{Q_F}$ in equation (10) has full column rank.

Thus, we arrive at the counterpart of Proposition 1 for the fuzzy RD case:

**Proposition 1F** Under Assumptions DB, 1F, 2F, 3F, 4 and 5, the distributions of $X^*$ conditional on $D$ and $u$ are identified.

It is worth noting as a theoretical point that identification is possible in the absence of Assumptions 2F, 4 and 5, provided that the researcher has exact knowledge of what $U_u$ and $L_u$ are. In this case, $L_{X^*|D=d}$ and $U_{X^*|D=d}$ can be recovered using this knowledge, and the probability masses are identified analogously if the full rank condition is satisfied. In practice, this observation has little practical value since researchers rarely–if at all–know the true values of $U_u$ and $L_u$. Thus, results may depend crucially on the imposed support, and the act of imposing support should be carried out with caution in empirical studies.

A related point is that identification can be obtained with only the independence assumption (Assumption 1 in the sharp case and Assumption 1F in the fuzzy case) if the econometrician has explicit knowledge of the marginal distribution of $X^*$, say from a validation sample.[13] This is because, as it is easy to show, the marginal distribution of $u$ is identified from the marginal distribution of $X$ and $X^*$ by an overidentified linear system of equations. It follows that the distribution of $X^*$ conditional on $D^*$ in the sharp case or conditional on $D$ in the fuzzy case is identified from the observed $X|D^*$ distribution in the sharp case or $X|D$ distribution in the fuzzy case together with the identified $u$ distribution. In practice, however, it is unlikely that an econometrician can obtain the marginal distribution of $X^*$ in the case of a transfer program even if s/he has access to administrative earnings data. First of all, commonly used administrative earnings records are of *quarterly* frequency, but program eligibility is usually based on *monthly* income. Second, the income used for determining program eligibility is typically earnings after certain deductions (child care or work related expenses, for example) plus unearned income. In that sense, the administrative earnings records are also a noisy version of the income for program eligibility determination, not to mention the fact that they may not perfectly measure true earnings either (e.g. Abowd and Stinson (2013)). That said, the possibility of obtaining the marginal distribution of $X^*$ for other applications should not be overlooked.

Finally, one might question the implicit assumption that benefit receipt $D$ is accurately measured when discussing the measurement error in $X$.[14] For example, errors in reporting program participation status in means-tested transfer programs have been documented in validation studies of survey data. Marquis and Moore (1990) reports that the AFDC under-reporting rate (i.e. those who did not report AFDC receipt among all who received the benefit) in the 1984 SIPP panel could be as high as 50%. The problem with under-reporting Medicaid coverage is also present but appears to be less severe–Card et al. (2004) estimate that the overall error rate in the 1990-93 SIPP Medicaid status is 15% for the state of California.

Under-reporting, however, does not pose a threat to the identification of the $X^*$ distribution, provided that those with $D = 1$ indeed received benefits and were therefore eligible. It follows that the support of $X^*$ conditional on $D$ and $u$ are identified correctly and that probability masses can be recovered, as long as

---

[13]This is the starting point of the approach taken by Davezies and Le Barbanchon (2014).

[14]For formal studies on the identification and estimation of regression models/treatment effects under misclassification of the treatment variable, see the seminal papers by Mahajan (2006) and Lewbel (2007).

Assumption 1F holds. It will be problematic, however, if those who do not take part in the program report participation. Fortunately, the rate of false-positive reporting associated with transfer programs at least is very small empirically–around 0.2% in the Marquis and Moore (1990) study and 1.5% in Card et al. (2004). This suggests that the reporting error in $D$ does not pose a big threat to the procedure above when applying an RD design using benefit discontinuities at the eligibility cutoff. Further, trimming procedures can be undertaken to correct for the false-positive reporting problem, which is further discussed in section 3.5.

## 3.3 Identification of the Conditional Expectation Functions and the RD Treatment Effect

In this section, we show that the conditional expectation function $E[Y|X^*]$ in a sharp design and $E[Y|X^*]$ and $E[D|X^*]$ in a one-sided fuzzy design can be identified under conditional versions of Assumptions 1, 2 and 3. In essence, these assumptions allow the performance of the deconvolution exercise detailed in section 3 for each value of $Y$, the outcome variable. Once we obtain the distribution of $X^*$ conditional on each value of $Y$, we apply Bayes' Theorem to recover $E[Y|X^*]$ in the sharp RD case and both $E[D|X^*]$ and $E[Y|X^*]$ in the fuzzy case. Finally, as with any RD design with a discrete assignment variable, we can parametrically extrapolate these conditional expectation functions to recover the RD treatment effect.[15] For ease of exposition, we focus on the case with a binary $Y$ and discuss identification with a general $Y$ distribution at the end of this subsection.

In the sharp RD model (1), the treatment effect is $\delta_{sharp} = \text{extp}_{c \uparrow 0} E[Y|X^* = c] - E[Y|X^* = 0]$ for discrete $X^*$. The first term is the left intercept of the $E[Y|X^*]$ function parametrically extrapolated using $E[Y|X^* = x^*]$ for $x^* < 0$, and we use "extp" to denote this extrapolation operator (analogous to the limit operator in the continuous $X^*$ case). The second term $E[Y|X^* = 0]$ is directly observed from the data. In order to identify $\delta_{sharp}$, we need to identify the conditional expectation function $E[Y|X^*]$, for which we propose the assumptions below. These assumptions imply that Assumptions 1, 2 and 3 hold conditional on $Y$:

**Assumption 1Y (Nondifferential Measurement Error).** $u \perp\!\!\!\perp X^*, Y$.[16]

**Assumption 2Y (Threshold Support).** $-1, 0 \in \text{support}_{X^*|Y=y}$ for each $y = 0, 1$.

---

[15]As pointed out by Lee and Card (2008), there may be a misspecification error in the parametric extrapolation of $E[Y|X^* = 0, D = 1]$. In this chapter, we abstract away from this error.

[16]Nondifferential measurement error is a commonly made assumption in the literature (see Carroll et al. (2006) for reference). As with Assumption 1F, a weaker version of Assumption 1Y, $X^* \perp\!\!\!\perp u$ conditional on $Y$, also delivers the following identification results. However, we adopt Assumption 1Y for its simplicity in economic interpretation. This is also the case for Assumption 1FY for exactly the same reason.

In order to to state Assumption 3 conditional on $Y$, note that Assumption 1Y and Assumption 2Y allow the formulation of the conditional counterparts of (7), $\mathbf{Q}_Y \mathbf{p}_Y = \mathbf{b}_Y$ for $Y = 0, 1$, where $\mathbf{Q}_Y$ and $\mathbf{p}_Y$ consist of probability masses of $q_j^1, q_j^0$, $p_k^1$ and $p_k^0$ (for the conditional distributions of $X$ and $X^*$ on $D^* = 1$ and $0$ respectively) conditional on $Y$.

**Assumption 3Y (Full Rank).** The matrix $\mathbf{Q}_Y$ is of full rank for $Y = 0, 1$.

**Proposition 2** Under Assumptions DB, 1Y, 2Y and 3Y, the conditional expectation function $E[Y|X^*]$ is identified for model (1).

*Proof.* Assumption 1Y implies that $X^* \perp\!\!\!\perp u$ conditional on $Y$. Therefore, the distribution of $X^*$ is identified from the observed joint distribution of $(X, D^*)$ conditional on each value of $Y$ by Proposition 1. That is, we can obtain the $X^*$ distribution conditional on $Y$, $\Pr(X^* = x^*|Y = y)$ for all $x^*$ and $y = 0, 1$. Consequently, $E[Y|X^* = x^*]$ is recovered by Bayes' Theorem since the marginal distribution of $Y$ is observed in the data. In the binary case,

$$E[Y|X^* = x^*] = \frac{\Pr(X^* = x^*|Y = 1)\Pr(Y = 1)}{\sum_y \Pr(X^* = x^*|Y = y)\Pr(Y = y)} \quad (11)$$

□

Once we pin down $E[Y|X^* = x^*]$, $\delta_{sharp}$ is subsequently identified by parametric extrapolation. Identification of the RD treatment effect parameter is obtained analogously in a fuzzy RD setting, with the only difference being the need to recover the first stage relationship $E[D|X^*]$. Consider formally the fuzzy RD model where the assignment variable is measured with error:

$$\begin{aligned} Y &= y(D, X^*, \varepsilon) \\ X &= X^* + u \end{aligned} \quad (12)$$

where the outcome $Y$ depends on program participation $D$. Under independence (Hahn et al. (2001)) or smoothness assumptions (DiNardo and Lee (2011)), the ratio

$$\delta_{fuzzy} = \frac{\operatorname*{extp}\limits_{c \uparrow 0} E[Y|X^* = c] - E[Y|X^* = 0]}{\operatorname*{extp}\limits_{c \uparrow 0} E[D|X^* = c] - E[D|X^* = 0]},$$

is the average treatment effect of $D$ on $Y$ for the "complier" population that takes up benefit when eligible. $\delta_{fuzzy}$ is the RD treatment effect to be identified in model (12), and the identification strategy is analogous to that in the sharp case: First identify the $X^*$ distribution conditional on $D$ and $Y$, then apply Bayes' Theorem to recover the conditional expectations of $Y$ and $D$ on $X^*$. Again, assumptions underpinning Proposition 1F are extended to hold conditional on $Y$:

**Assumption 1FY (Strong Independence and Nondifferential Measurement Error: Fuzzy).** $u \perp\!\!\!\perp X^*, D, Y$.

**Assumption 2FY (Threshold Support: Fuzzy).** $-1 \in \text{support}_{X^*|D=1,Y=y}$ and $0 \in \text{support}_{X^*|Y=y}$ for each $y = 0, 1$.

As in the sharp case, in order to state Assumption 3F conditional on $Y$, note that Assumption 1FY and Assumption 2FY allow the formulation of the conditional counterparts of (10), $\mathbf{Q_{FY}} \mathbf{p_{FY}} = \mathbf{b_{FY}}$ for $Y = 0, 1$, where $\mathbf{Q_{FY}}$ and $\mathbf{p_{FY}}$ consist of probability masses of $q_j^1, q_j^0, p_k^1$ and $p_k^0$ (for the conditional distributions of $X$ and $X^*$ on $D = 1$ and $0$ respectively) conditional on $Y$.

**Assumption 3FY (Full Rank: Fuzzy).** The matrix $\mathbf{Q_{FY}}$ is of full rank for $Y = 0, 1$.

**Proposition 2F** Under Assumptions DB, 1FY, 2FY, 3FY, 4 and 5, the conditional expectation functions $E[Y|X^*]$ and $E[D|X^*]$ are identified for model (12).

*Proof.* Analogous to arguments in the previous section, Assumptions DB, 1FY, 2FY, 3FY, 4 and 5 are sufficient conditions for identifying the $X^*$ distribution conditional on $D = d$ and $Y = y$ for $d, y = 0, 1$ from that of $(X, D)|Y$. It follows that $\Pr(X^* = x^*|D = d)$ and $\Pr(X^* = x^*|Y = y)$ for $d, y = 0, 1$ are identified because $\Pr(Y = y|D = d)$ and $\Pr(D = d|Y = y)$ are observed in the data:

$$\Pr(X^* = x^*|D = d) = \sum_y \Pr(X^* = x^*|D = d, Y = y) \Pr(Y = y|D = d) \tag{13}$$

$$\Pr(X^* = x^*|Y = y) = \sum_d \Pr(X^* = x^*|D = d, Y = y) \Pr(D = d|Y = y) \tag{14}$$

Consequently, $E[D|X^* = x^*]$ and $E[Y|X^* = x^*]$ are recovered by an application of the Bayes' Theorem

$$E[D|X^* = x^*] = \frac{\Pr(X^* = x^*|D = 1) \Pr(D = 1)}{\sum_d \Pr(X^* = x^*|D = d) \Pr(D = d)} \tag{15}$$

$$E[Y|X^* = x^*] = \frac{\Pr(X^* = x^*|Y = 1)\Pr(Y = 1)}{\sum_y \Pr(X^* = x^*|Y = y)\Pr(Y = y)}. \tag{16}$$

□

Propositions 2 and 2F can be easily generalized from the formulation with a binary $Y$. Let $F_Y$ denote the cumulative distribution function of $Y$, which is observed by the econometrician. We can still identify $\Pr(X^*|Y = y)$ for each value of $y$ and subsequently identify $E[Y|X^* = x^*]$ in both the sharp and fuzzy design using

$$E[Y|X^* = x^*] = \frac{\int y \Pr(X^* = x^*|Y = y) dF_Y(y)}{\int \Pr(X^* = x^*|Y = y) dF_Y(y)}. \tag{17}$$

## 3.4 Estimators of the Assignment Variable Distribution and the RD Treatment Effect

In this section, we propose estimators for the assignment variable distribution and the RD treatment effect in the discrete and bounded case. As in identification, estimation of the $X^*$ distribution follows two steps: estimation of its support and estimation of the probability masses at each point in its support. Estimation of support follows Equations (8) and (9) with the population quantities replaced by sample analogs. We abstract away from the sampling error of support and simply assume that the sample is large enough to reveal the true support of the distribution. We present the case in the sharp design setting where we omit the $F$ subscript for notational convenience. Results in the fuzzy case follow by replacing $D^*$ by $D$.

Given the specification of probability model with independent measurement error, the likelihood function can be explicitly written out by using the $p_k^1$'s, $p_k^0$'s, $m_l$'s and the marginal probability $r = \Pr(D^* = 1)$. Formally, the likelihood for the joint distribution $(X, D^*)$ is

$$\begin{aligned} L(X_i, D_i^*) &= L(X_i|D_i^*) L(D_i^*) \\ &= \{(\sum_k p_{X_i-k}^1 m_k) r\}^{D_i^*} \{(\sum_k p_{X_i-k}^0 m_k)(1-r)\}^{1-D_i^*} \end{aligned} \tag{18}$$

Researchers can directly estimate (18) and the resulting estimators are efficient provided that the parameters are in the interior of the parameter space, i.e. strictly between zero and one. However, the analytical solutions to maximizing the log likelihood do not appear to exist, and numerically optimizing (18) may become computationally intensive as the number of points in the support of $X^*$ and $u$ increases.

An alternative strategy relies on the identification equation (7), which fits nicely into a standard minimum distance framework $f(\mathbf{q}, \mathbf{p}) = \mathbf{Qp} - \mathbf{b} = 0$ (Kodde et al. (1990)) from which an estimator of $\mathbf{p}$ can be obtained easily.[17] Because of the linearity in (7), the parameter vector of interest **p** can be estimated analytically once an estimator of **q** are obtained. Estimation follows the following steps:

1. Obtain the estimators $\hat{q}_j^1 = \frac{\sum_i 1_{[X_i=j]} \cdot 1_{[D_i=1]}}{\sum_i 1_{[D_i=1]}}$, $\hat{q}_j^0 = \frac{\sum_i 1_{[X_i=j]} \cdot 1_{[D_i=0]}}{\sum_i 1_{[D_i=0]}}$ and $\hat{r} = \frac{1}{N}\sum 1_{[D_i=1]}$ ($N$ denotes the sample size), as well as $\hat{\Omega}$, which is a consistent estimator for the asymptotic variance-covariance matrix $\Omega$ of $\hat{q}$, (that is, $\sqrt{N}(\hat{q}-q) \Rightarrow N(0,\Omega)$). Since $X|D^* = d$ follows a multinomial distribution for each $d$, $\hat{\Omega}$ is a block-diagonal matrix $\hat{\Omega} = \begin{bmatrix} \hat{\Omega}^{11} & \hat{\Omega}^{10} \\ \hat{\Omega}^{01} & \hat{\Omega}^{00} \end{bmatrix}$ where $\hat{\Omega}^{01} = \hat{\Omega}^{10} = 0$, and

$$\hat{\Omega}_{ij}^{dd} = \begin{cases} (1-\hat{q}_i^d)\hat{q}_i^d/(d\hat{r}+(1-d)(1-\hat{r})) & \text{if } i = j \\ -\hat{q}_i^d\hat{q}_j^d/(d\hat{r}+(1-d)(1-\hat{r})) & \text{if } i \neq j \end{cases}$$

2. Form the estimator of the **Q** matrix under perfect compliance in (7), $\hat{\mathbf{Q}}$, by replacing the $q_j^1$ and $q_j^0$ in **Q** with their estimators;

3. Derive a consistent estimator of **p**: $\hat{\mathbf{p}} = \arg\min_{\mathbf{p}} f(\hat{\mathbf{q}},\mathbf{p})'f(\hat{\mathbf{q}},\mathbf{p}) = (\hat{\mathbf{Q}}'\hat{\mathbf{Q}})^{-1}(\hat{\mathbf{Q}}'\mathbf{b})$;

4. Compute the optimal weighting matrix $\widehat{\mathbf{W}} = (\widehat{\nabla_{\mathbf{q}} f}\hat{\Omega}\widehat{\nabla_{\mathbf{q}} f}')^{-1}$ where $\hat{\Omega}$ is a consistent estimator for the variance-covariance matrix of the **q** derived in step 1.[18] $\widehat{\nabla_{\mathbf{q}} f}$ is a consistent estimator for $\nabla_{\mathbf{q}} f$, the Jacobian of $f$ with respect to **q**. Because $\nabla_{\mathbf{q}} f$ depends on **p**, step 3 was necessary for first obtaining a consistent estimate of **p**. It turns out that $f$ is also linear in **q**, and hence $\widehat{\nabla_{\mathbf{q}} f}$ can be computed analytically;

5. Arrive at the optimal estimator of **p**: $\hat{\mathbf{p}}_{opt} = (\hat{\mathbf{Q}}'\widehat{\mathbf{W}}\hat{\mathbf{Q}})^{-1}(\hat{\mathbf{Q}}'\widehat{\mathbf{W}}\mathbf{b})$.

Provided that the true parameter lies in the interior of the parameter space:

**Assumption 6 (Interior).** $\mathbf{p} \in (0,1)^K$ where $K$ is the length of **p**

The derivation of the asymptotic distribution of $\hat{\mathbf{p}}_{opt}$ is standard. Specifically,

**Proposition 3** Under Assumptions DB, 1, 2, 3 and 6 for the perfect compliance case, $\sqrt{N}(\hat{\mathbf{p}}_{opt}-\mathbf{p}) \Rightarrow \mathcal{N}(0,\mathbf{Q}'(\nabla_{\mathbf{q}} f \Omega \nabla_{\mathbf{q}} f')^{-1}\mathbf{Q})$ where $\Omega$ is the asymptotic variance-covariance matrix of **q**, and **Q** along with $\nabla_{\mathbf{q}} f = \nabla_{\mathbf{q}}(\mathbf{Q}\mathbf{p}-\mathbf{b})$ are specified in Equation (7).

---

[17]Analogous to **p**, which contains the parameters $p_k^0$'s and $p_k^1$'s, **q** is the vector that contains all the parameters $q_j^0$'s and $q_j^1$'s.

[18]It turns out that $\nabla_{\mathbf{q}} f \Omega \nabla_{\mathbf{q}} f'$ is singular and is analogous to the case covered in a recent paper by Satchachai and Schmidt (2008) where there are too many restrictions. The study advised against using the generalized inverse, which is confirmed by our own numerical investigation. Instead, they propose dropping one or more restrictions, but stated that the problem of which restrictions to drop has not yet been solved. All the empirical results are based on the last row of the **Q** matrix being dropped (dropping other rows had little impact on the empirical results).

Analogously, for the imperfect compliance case, we have

**Proposition 3F** Under Assumptions DB, 1F, 2F, 3F, 4, 5 and 6, $\sqrt{N}(\hat{\mathbf{p}}_{Fopt} - \mathbf{p_F}) \Rightarrow \mathcal{N}(\mathbf{0}, \mathbf{Q}'_\mathbf{F}(\nabla_{\mathbf{q_F}} f_\mathbf{F} \Omega_\mathbf{F} \nabla_{\mathbf{q_F}} f'_\mathbf{F})^{-1}\mathbf{Q})$ where $\Omega_\mathbf{F}$ is the asymptotic variance-covariance matrix of $\mathbf{q_F}$, and $\mathbf{Q_F}$ along with $\nabla_{\mathbf{q_F}} f_\mathbf{F} = \nabla_{\mathbf{q_F}} (\mathbf{Q_F p_F} - \mathbf{b_F})$ are specified in Equation (10).

The main conclusion of Kodde et al. (1990) shows that $\hat{\mathbf{p}}_{opt}$ is efficient – it has the same asymptotic variance as the maximum likelihood estimator – if $\mathbf{p}$ is exactly or over-identified by $f(\mathbf{q}, \mathbf{p}) = 0$, and $\hat{\mathbf{q}}$ is the maximum likelihood estimator. Since both conditions are satisfied in our setup, we have obtained a computationally inexpensive estimator without sacrificing efficiency. Also note that the assumptions can be jointly tested by the overidentifying restrictions as is standard for minimum distance estimators. In particular, the test statistic $N \cdot (f(\hat{\mathbf{q}}, \hat{\mathbf{p}}_{opt})' \hat{\mathbf{W}} f(\hat{\mathbf{q}}, \hat{\mathbf{p}}_{opt}))$ in the sharp case or $N \cdot (f(\hat{\mathbf{q}}_\mathbf{F}, \hat{\mathbf{p}}_{Fopt})' \hat{\mathbf{W}}_\mathbf{F} f(\hat{\mathbf{q}}_\mathbf{F}, \hat{\mathbf{p}}_{Fopt}))$ in the fuzzy case follows a $\chi^2$-distribution with degrees of freedom equal to $K_u$, the number of points in the support of the measurement error, when assumptions in Proposition 3 or Proposition 3F are satisfied.[19]

A concern arises because the optimal estimators of $p_k^1$ and $p_k^0$ may never sum to one by following the procedure above. Therefore, we need to modify the estimation strategy and impose this constraint, as oppose to simply minimizing the distance between the sums and one. The following modifications incorporate the matrix constraints $\mathbf{Rp} = \mathbf{c}$, where $\mathbf{c} = \begin{bmatrix} 1 \\ 1 \end{bmatrix}$ and $\mathbf{R} = \begin{bmatrix} 1 & \cdots & 1 & 0 & \cdots & 0 \\ 0 & \cdots & 0 & 1 & \cdots & 1 \end{bmatrix}$ that summarize the restriction that the $p_k^1$ and $p_k^0$ sum to one. In step 3, the consistent estimator is instead $\hat{\mathbf{p}} = (\hat{\mathbf{Q}}'\hat{\mathbf{Q}})^{-1}\mathbf{R}'\{\mathbf{R}(\hat{\mathbf{Q}}'\hat{\mathbf{Q}})^{-1}\mathbf{R}'\}^{-1}\mathbf{c}$[20], and in step 5, $\hat{\mathbf{p}}_{opt} = (\hat{\mathbf{Q}}'\hat{\mathbf{W}}\hat{\mathbf{Q}})^{-1}\mathbf{R}'\{\mathbf{R}(\hat{\mathbf{Q}}'\hat{\mathbf{W}}\hat{\mathbf{Q}})^{-1}\mathbf{R}'\}^{-1}\mathbf{c}$, and finally the asymptotic variance of $\hat{\mathbf{p}}_{opt}$ is given by $\mathbf{T}((\mathbf{QT})'\mathbf{W}(\mathbf{QT}))^{-1}\mathbf{T}'$ where $\mathbf{T}$ is a matrix whose columns are the first $K-2$ eigenvectors of the projection matrix $\mathbf{I} - \mathbf{R}'(\mathbf{RR}')^{-1}\mathbf{R}$. The computation for $\hat{\mathbf{W}}$ is unaltered by the imposition of the linear constraints $\mathbf{Rp} = \mathbf{c}$.

In order to construct the asymptotic distribution of the RD treatment effect estimators, we need to estimate the variance covariance matrix of $E[Y|X^* = x^*]$ in the sharp RD case and $E[D|X^* = x^*]$ and $E[Y|X^* = x^*]$ in the fuzzy RD case for each $x^*$ in the support of $X^*$. We refer to (13), (14), (15) and (16), which show that these two entities are differentiable functions of $\Pr(X^* = x^*|D = d, Y = y)$, $\Pr(D = d|Y = y)$ and $\Pr(Y = 1)$ for $d, y = 0, 1$. The delta method can be directly applied, where the Jacobian of the transformations is derived analytically. A general expression of the RD treatment effect estimator cannot be obtained because it depends on the the functional form of $E[Y|X^*]$ and $E[D|X^*]$ which varies from application to application. Therefore, we cannot provide the asymptotic distribution for $\hat{\delta}_{sharp}$ and $\hat{\delta}_{fuzzy}$ in the general case.

---

[19]Identification in section 3.2 is established under the symmetry assumption (Assumption 5), and the model may not be identified in its absence. Therefore, this overidentification test may not have power against the violation of Assumption 5.

[20]For a clear exposition of this standard result, see the "makecns" entry in Stata (2010).

Finally, we note that the restriction of to a binary $Y$ can be relaxed. First, estimation proceeds in exactly the same way when $Y$ takes on a finite number of values. Second, our estimator can be further generalized to the case where $Y$ is continuously distributed.[21] As a special case of (17),

$$E[Y|X^* = k] = \int y \frac{\Pr(X^* = k|Y = y) f_Y(y)}{\Pr(X^* = k)} dy.$$

The idea for the next step is to approximate the integral by the sum of the terms evaluated on grids of $y$. We can estimate $\Pr(X^* = k)$ as before and estimate $f_Y(y)$ with a standard kernel density estimator. For $\Pr(X^* = k|Y = y)$, take a sequence of bandwidths denoted by $\{h_n\}_{n=1,2,...}$ satisfying $h_n \to 0$ and $nh_n \to \infty$ as $n \to \infty$. For each $n$, we can identify the supports of $X^*$ and $u$ condition on $Y \in [y - h_n, y + h_n]$. Asymptotically this will give the true support of $X^*$ and $u$ condition on $Y = y$ under mild conditions (e.g., the support of $X^*$ given $y$ only changes on a discrete set of $y$). Meanwhile, it is standard to estimate the distribution of $X$ given $D^* = d$ and $Y = y$ for $d = 0, 1$, despite the fact that $D^*$ is discrete and $Y$ is continuous.[22] This, along with the information on the supports of $X^*$ and $u$, allows us to identify $\Pr(X^*|Y = y)$ as before. This approach has two drawbacks. First, in order to closely approximate the integral, the procedure is computationally burdensome. Second, if the support of $X$ is wide, the procedure will require a large sample size for the estimates to be reliable. Therefore, in the numerical and empirical illustrations below, we adhere to a binary outcome variable.

## 3.5 Potential Issues in Practical Implementation

There are several issues in implementing the procedure described in section 3.4. First of all, in order to have realistic support for the true assignment variable, the maximum value of the observed assignment variable needs to be significantly larger for the $D = 0$ group than for the $D = 1$ group, since the difference of the two is the upper end point in the true assignment variable distribution. Also, the left tail of the observed assignment variable distribution may need to be significantly longer that of the right tail of the $D = 1$ group since the difference in the lengths is the lower bound of the true assignment variable distribution (following

---

[21] We thank a referee for suggesting this extension.
[22] See Li and Racine (2007) for details in nonparametric density estimation methods with mixed covariates.

Assumption 5). Since symmetry is a functional form assumption, which may not hold when the assignment variable is in levels (e.g. in the case of income), a transformation of the observed assignment variable may be needed. In practice, a Box-Cox type transformation is recommended and practitioners may experiment with various transformation parameters. The overidentification test mentioned in the previous section can be used to help decide which transformation parameters to use.

A related point, as mentioned at the end of section 3.2, is that someone with very large observed $X$ and not program eligible may actually report program participation ($D = 1$) by mistake. If this is the case, the supports will not be correctly identified, and using a Box-Cox transformation will not be sufficient to correct the problem. A trimming procedure should be adopted in practice where outliers in both the left and right tails of the $X|D = 1$ and $X|D = 0$ populations may be dropped. As with the case of transformation parameters, we recommend trying several trimming percentages and examining the sensitivity of the empirical results. Finally, a quadratic programming routine with inequality constraints can be used in practice to guarantee nonnegativity of the probability masses.

### 3.6 Illustration with a Simple Numerical Example

In this section, we illustrate the proposed estimation procedure in section 3.4 with a simple numerical example. We focus on the more complicated fuzzy case and show that the true first stage and outcome functions as well as the $X^*$ distribution can indeed be recovered when the assumptions in Proposition 2F and 3F are met. In the baseline example, we generate $X^*$ following a uniform distribution on the set of integers from -10 to 10. $u$ follows a uniform distribution between -3 and 3 and is therefore symmetric in its support (Assumption 5). The true first stage relationship is given by

$$E[D|X^*] = \Pr(D = 1|X^*) = (\alpha_{D^*X^*}X^* + \alpha_{D^*})1_{[X^*<0]} = \alpha_{D^*}D^* + \alpha_{D^*X^*}D^*X^* \tag{19}$$

which reflects the one-sided fuzzy assumption (Assumption 4), and the size of the first stage discontinuity is $\alpha_{D^*}$. The outcome response function is given by the simple constant treatment effect specification

$$E[Y|X^*,D] = \Pr(Y = 1|X^*,D) = \delta_0 + \delta_1 X^* + \delta_{fuzzy}D \tag{20}$$

where the treatment effect to be identified is $\delta_{fuzzy}$. Note that (19) and (20) together imply that the second stage of $Y$ versus $X^*$ is

$$E[Y|X^*] = \Pr(Y = 1|X^*) = \beta_0 + \beta_{D^*}D^* + \beta_1 X^* + \beta_{D^*X^*}D^*X^* \tag{21}$$

where $\beta_0 = \delta_0$, $\beta_{D^*} = \alpha_{D^*}\delta_{fuzzy}$, $\beta_1 = \delta_1$ and $\beta_{D^*X^*} = \alpha_{D^*X^*} \cdot \delta_{fuzzy}$.

Figures 2 and 3 present graphical results based on a sample of 25,000 generated observations for the parameter values $\alpha_{D^*X^*} = -0.01$, $\alpha_{D^*} = 0.8$, $\delta_0 = 0.15$, $\delta_1 = -0.01$, and $\delta_{fuzzy} = 0.6$, with the implied coefficients in (21) being $\beta_0 = 0.15$, $\beta_{D^*} = 0.48$, $\beta_1 = -0.01$ and $\beta_{D^*X^*} = -0.006$. We choose $N = 25,000$ because it is about the average sample size in the relevant studies – 45,722 in Hullegie and Klein (2010), 34,949 in Koch (2013), 11,541 in Schanzenbach (2009) and 2,954 in De La Mata (2012). The top and bottom panels in Figure 2 plot the *observed* first and second stage, i.e. $E[D|X]$ and $E[Y|X]$, respectively. Note that there is no visible discontinuity at the thresholds, and the estimated first-stage and outcome discontinuities based on these observed relationships cannot identify the true parameter values of $\alpha_{D^*}$ and $\beta_{D^*}$, which are 0.8 and 0.48 respectively.

Figure 3 plots the *estimated* first and second stage based on procedures developed in section 3.4 against the actual (19) and (20) specified with the parameter values above. As is evident from the graphs, the proposed procedures can correctly recover the true first-stage and outcome relationships of the underlying RD design. $\hat{\delta}_{fuzzy}$, the RD treatment effect parameter, is obtained by fitting another linear minimum distance procedure on the estimated $E[D|X^*]$ and $E[Y|X^*]$ (as well as their estimated variance-covariance matrices) with the parametric restrictions (19) and (21). In 1,000 repeated samples, the average point estimate for $\alpha_{D^*}$ is 0.75 (true parameter value is 0.8), the average standard error is 0.063, and the coverage rate of the 95% confidence interval is 97%; the average point estimate for $\beta_{D^*}$ is 0.48 (true parameter value is 0.48), the average standard error is 0.075 and the coverage rate of the 95% confidence interval is 98%.

To gauge the performances of the estimators in adverse settings, we test their sensitivity to the violation of symmetry and to larger supports in $X^*$ and $u$ relative to the sample size. Unfortunately, the performance of the estimators deteriorates when the symmetry assumption is violated. For example, when $u$ is supported on the integers in $[-4,3]$ but the lower bound of its support is erroneously assumed to be $-3$, the average point estimates for $\alpha_{D^*}$ and $\beta_{D^*}$ are 0.66 and 0.39, and the coverage rates of the 95% confidence interval are 0.56 and 0.74, respectively. Admittedly, this is a limitation of the proposed method. On the other hand, the behavior of the estimators is more robust as support$_X$ becomes larger relative to the sample size – as the set of points in support$_X$ increases from 27 in the numerical example above to 37, the coverage rates are still around 95%. The coverage rates fall to around 80% and 55% when |support$_X$| is 47 and 67, respectively.

As mentioned above, the proposed procedure can also be used to estimate the discontinuity in the density of $X^*$ at the eligibility threshold, which is often used to evaluate the validity of the RD design but may in addition shed light on economically interesting quantities (as per Saez (2010)). We perform another numerical exercise to assess the ability of the estimation method to detect nonsmoothness in the $X^*$ distribution. In particular, we consider two alternative specifications: 1) we consider the specification above (i.e. that used for Figures 2 and 3), for which there is no discontinuity in the $X^*$ distribution at the eligibility threshold; 2) $X^*$ is still supported on the set of integers from -10 to 10 but with a discontinuity at the eligibility threshold: $\Pr(X^* = i) = 0.06$ for each $i < 0$ and $\Pr(X^* = i) = 0.036$ for $i \geq 0$. Figures 4 and 5 present the observed $X$ and *estimated* $X^*$ distribution for cases 1) and 2) respectively. Note that there is no obvious discontinuity in the observed $X$ distribution at the eligibility threshold in case 2) (top panel of Figure 5)–the measurement error has simply smoothed it over. This lack of observed threshold discontinuity illustrates again the problematic nature of using the observed assignment variable $X$ for RD analyses. In both cases, we test for the threshold discontinuity by fitting a linear minimum distance procedure on the estimated $X^*$ distribution with the restriction

$$\Pr(X^* = x^*) = \gamma_0 + \gamma_{D^*} D^* + \gamma_1 X^* + \gamma_{D^* X^*} D^* X^*$$

Let $\gamma_{D^*}^{1)}$ and $\gamma_{D^*}^{2)}$ be the coefficients of $\gamma_{D^*}$ in cases 1) and 2) respectively, and based on the specifications above, $\gamma_{D^*}^{1)} = 0$ and $\gamma_{D^*}^{2)} = 0.023$. In 10,000 repeated samples with 25,000 observations: the average value of $\gamma_{D^*}^{1)}$ is -0.003, and the coverage rate of the 95% confidence interval is 89%; the average value of the $\gamma_{D^*}^{2)}$ estimates is 0.031, and the coverage rate of the 95% confidence interval is 87%. The coverage rates for the density discontinuity confidence intervals improve to 90% or higher as the sample size exceeds 50,000. Overall, this simple numerical example verifies that the true assignment variable distribution and the RD treatment effect parameter can indeed be recovered using the proposed method.

## 4 Continuous Assignment Variable and Measurement Error

In this section, we study the identification in an RD design when $X^*$ and $u$ are continuously distributed and discuss the sharp and fuzzy cases separately in the two subsections below. Before we proceed, we introduce the analog of Assumption 2 in the continuous case, a standard assumption in the RD literature.

**Assumption 2C (Positive Density at Threshold).** Let the p.d.f. of $X^*$ be $f_{X^*}$. There exists $a > 0$ such that $f_{X^*}(x^*) > 0$ for $x^* \in (-a, a)$.

Assumption 2C ensures that the conditional expectation functions are well-defined in the sharp and fuzzy RD estimands:

$$\delta_{sharp}^c = \lim_{c \uparrow 0} E[Y|X^* = c] - \lim_{c \downarrow 0} E[Y|X^* = c]$$

$$\delta_{fuzzy}^c = \frac{\lim_{c \uparrow 0} E[Y|X^* = c] - \lim_{c \downarrow 0} E[Y|X^* = c]}{\lim_{c \uparrow 0} E[D|X^* = c] - \lim_{c \downarrow 0} E[D|X^* = c]}$$

## 4.1 Identification under Perfect Compliance

We focus on the case of perfect compliance in this subsection and consider three distinct approaches. First, we obtain semiparametric identification of the $X^*$ distribution and the RD treatment effect by imposing normality on the distribution of $u$. Second, we propose a nonparametric identification-at-infinity strategy for the RD treatment effect by restricting the tail behavior of the $X^*$ distribution. Finally, we discuss how we can apply the nonparametric simulation extrapolation (SIMEX) method of Carroll et al. (1999), which requires knowledge of the measurement error variance, to recover the RD treatment effect.

**Approach 1.** In the first approach, we assume that $u$ follows a normal distribution with mean zero and an unknown variance $\sigma^2$. Along with the classical measurement error assumption, this assumption allows the distribution $X^*$ to be identified. Provided that the measurement error is nondifferential and that the density of $X^*$ is positive in a neighborhood of zero, the RD treatment effect is also identified.

**Assumption 7 (Normality).** $u \sim \phi(0, \sigma^2)$.

**Proposition 4** (a) Under Assumptions 1 and 7, the distributions of $X^*$ and $u$ are identified. (b) Under Assumptions 1Y, 2C and 7, $\delta_{sharp}^c$ is identified.

Proposition 4(a) is a corollary of Theorem 2.1 of Schwarz and Bellegem (2010), who prove the identification of $\sigma$ and the distribution of $X^*$ from the joint distribution of $(X, D^*)$. Intuitively, let $f_a$ and $f_b$ be two candidate distributions for $X^*|D^* = 1$, and let $\sigma_a$ and $\sigma_b$ ($\sigma_a < \sigma_b$) be two candidates for $\sigma$. Suppose $(f_a, \sigma_a)$ and $(f_b, \sigma_b)$ are observationally equivalent, i.e. $f_a * \phi(0, \sigma_a^2) = f_b * \phi(0, \sigma_b^2) = g$, where $g$ is the density function of $X|D^* = 1$. Then it follows from properties of normal random variables that $f_1 = f_2 * \phi(0, \sigma_2^2 - \sigma_1^2)$. A contradiction arises because $f_1$ is only supported on the negative part of the real line but $f_2 * \phi(0, \sigma_2^2 - \sigma_1^2)$ is supported on the entire real line. Hence, $\sigma_1 = \sigma_2$ and the continuous density of $X^*|D^* = 1$ on $x^* \in (-\infty, 0)$ is identified by the one-to-one correspondence between characteristic function and probability density $f_{X^*|D^*=1}(x^*) = \frac{1}{2\pi} \int_{-\infty}^{\infty} e^{-itx} \frac{\varphi_{X|D^*=1}(t)}{\varphi_u(t)} dt$, where $\varphi_A$ denotes the characteristic function of the random variable $A$ (note that $\varphi_u(t) = e^{-\frac{1}{2}\sigma^2 t^2}$, which appears in the denominator of the integrand, is nonzero for all $t$). The distribution of $X^*|D^* = 0$ is identified analogously. Since $\Pr(D^* = d)$ is observed for $d = 0, 1$, we can identify the unconditional $X^*$ distribution:

$$f_{X^*}(x^*) = \sum_{d=0}^{1} f_{X^*|D^*=d}(x^*) \Pr(D^* = d) \tag{22}$$

For Proposition 4(b), the idea of the proof is the same as that of Proposition 2. The identification of the $X^*$ density for each value of $Y$, $f_{X^*|Y=y}(y)$, is given by the combination of Assumption 1Y and part (a) of the proposition. We can then identify the conditional expectation function using the continuous analog of equation (17):

$$E[Y|X^* = x^*] = \frac{\int y f_{X^*|Y=y}(x^*) dF_Y(y)}{\int f_{X^*|Y=y}(x^*) dF_Y(y)}. \tag{23}$$

Assumption 2C guarantees that the denominator of (23) is nonzero and that $E[Y|X^* = x^*]$ is defined for $x^*$ in a neighborhood of zero. Taking the difference of the limit of $E[Y|X^* = x^*]$ across the threshold identifies $\delta_{sharp}^c$.

**Approach 2.** In the second approach, we show that normality of the measurement error can be weakened without sacrificing the identification of the RD treatment effect under perfect compliance. Specifically, an identification-at-infinity strategy can be applied with a regularity condition governing the tail of the measurement error.

**Proposition 5** If Assumption 2C holds, $f_{X^*}$ is continuous at 0, and the c.d.f. of $u$, $F_u$, satisfies

$$\lim_{x \to \infty} \frac{1 - F_u(x+v)}{1 - F_u(x)} = 0 \qquad \text{for all } v > 0, \tag{24}$$

$$\lim_{x \to -\infty} \frac{F_u(x-v)}{F_u(x)} = 0 \qquad \text{for all } v > 0, \tag{25}$$

then the RD treatment effect is identified:

$$\lim_{x \to \infty} E[Y|X = x, D^* = 1] - \lim_{x \to -\infty} E[Y|X = x, D^* = 0] = \delta_{sharp}^c. \qquad (26)$$

The proof of Proposition 5 is in the Appendix section D. Intuitively, when we see an observation with $X \gg 0$ and $D^* = 1$, it is because $X^*$ is close to zero or because $u$ is very large. Condition (24) states that the right tail of the measurement error distribution needs to be "light" enough and makes $u$ unlikely to be very large. Therefore, as $X$ becomes large in the treatment group, we end up with observations for which $X^*$ is just below zero. While the Laplace distribution violates (24) and (25), almost all of the common distributions whose tails are comparable to or lighter than the normal distribution satisfies these tail conditions.[23]

A visual illustration of the identification result in Proposition 5 is presented in Figure 6. We use the same conditional expectation function for $E[Y|X^*]$ as in section 3.6. $X^*$, $u$ and the error term in the outcome equation are all normally distributed. We split the simulated sample of 25,000 observations into two groups by the value of $D^*$. In the upper and lower panels, we show the raw scatter plot for the $D^* = 1$ and $D^* = 0$ groups, respectively. In each panel, we impose fit obtained from a simple local regression. As the $X$ becomes large in the $D^* = 1$ group and as $X$ becomes small in the $D^* = 0$ group, we can see that the averages of the $Y$'s (represented by the red local fit curves) approach the true intercepts of the $E[Y|X^*]$ function (represented by the black horizontal lines).

We now discuss the relationship between the identification-at-infinity strategy proposed here with recent papers by Yu (2012) and Yanagi (2014) that study the same question. Yu (2012) assumes that the measurement error variance shrinks to zero as the sample size increases and proposes using a trimmed sample to recover the RD treatment effect, in which observations with $X \geqslant 0$ in the $D^* = 1$ group and observations with $X < 0$ in the $D^* = 0$ group are dropped. Specifically, the RD treatment effect estimand is the difference between the left intercept of $E[Y|X = x, D^* = 1]$ and the right intercept of $E[Y|X = x, D^* = 0]$. As mentioned in section 2, Dividing the population into treatment and control and separately estimating the intercept in each is attractive, because it results in a much smaller bias than naively fitting $E[Y|X]$ with local regressions. Our approach builds on the idea of Yu (2012) in this regard. The difference is that our approach puts no restriction on the measurement error variance, and it may result in a smaller bias as a result. To see this, we illustrate the strategy of Yu (2012) in Figure 7, and the bias appears to be somewhat larger as compared to that in Figure 6.

---

[23] When $u$ follows a Laplace distribution, e.g., $f_u(v) = \frac{1}{2} e^{-v}$ for $v \in \mathbb{R}$, $\lim_{x \to \infty} E[Y|X = x, D^* = 1] = \int_0^\infty E[Y|X^* = -v] e^{-v} dv$, which is generally not equal to $\lim_{x^* \uparrow 0} E[Y|X^* = x^*]$.

To illustrate how our identification results relates to Yanagi (2014), first note that the proof of Proposition 5 can be carried over to the case where the measurement error $u$ has bounded support. When $\text{support}_u$ is bounded, the tail condition (25) is no longer needed. Furthermore, we can identify $\text{support}_u \equiv [\underline{u}, \bar{u}]$ and identify the RD treatment effect with

$$\lim_{x \to \bar{u}} E[Y|X = x, D^* = 1] - \lim_{x \to \underline{u}} E[Y|X = x, D^* = 0] = \delta_{sharp} \tag{27}$$

To illustrate (27), we present Figure 8, for which the underlying $u$ distribution is uniform as opposed to normal. In this figure, the local regression fit approaches the true RD intercept as $X$ approaches the boundary of $\text{support}_u$.

Assuming bounded measurement error $u$, Yanagi (2014) creatively applies a small error variance approximation (Chesher (1991)) to $E[Y|X^* = x, D^* = d]$ in a sharp design:

$$\begin{aligned}
E[Y|X^* = x, D^* = d] &= E[Y|X = x, D^* = d] \\
&\quad - \sigma^2 \left( \log^{(1)} f_{X|D}(x|d) \right) E^{(1)}[Y|X = x, D^* = d] \\
&\quad - \frac{\sigma^2}{2} E^{(2)}[Y|X = x, D^* = d] + o(\sigma^2)
\end{aligned} \tag{28}$$

where $\sigma^2$ again denotes the variance of $u$. For the approximation to work, however, the derivatives of the conditional expectation and density functions need to be estimated, and $\sigma$ needs to be small and known. In comparison, our "identification-at-boundary" strategy of (27) has the advantage of avoiding the derivatives in (28), which may be hard to estimate in practice. Furthermore, it does not place additional restriction on the measurement error distribution, nor does it require $\sigma$ to be known.

**Approach 3.** When $\sigma$ *is* known (e.g. from an external source), we can also apply the simulation-extrapolation, or SIMEX, strategy of Cook and Stefanski (1994) and Carroll et al. (1999) to recover the RD treatment effect. It is the final approach we propose in this section and is less restrictive than that of Yanagi (2014).[24] The idea behind the method is most simply illustrated in the linear case where the conditional expectation function $E[Y|X^* = x^*, D^* = 1]$ is equal to $\psi_0 + \psi_1 X^*$. When we regress $Y$ on $X^*$ within the $D^* = 1$ population, the least squares slope estimator converges to the expression with the well-known attenuation factor, $\psi_1 \cdot \frac{var(X^*|D^*=1)}{var(X^*|D^*=1)+\sigma^2}$. The attenuation problem worsens as $\sigma^2$ increases, and the SIMEX approach of Cook and Stefanski (1994) makes use of this observation, traces out the estimate as a function of the degree of measurement error contamination via simulation, and extrapolates this function to recover the true parameter $\psi_1$ as well as the target conditional expectation function. Formally, for any $\lambda > 0$, we can add additional noise $\tilde{u}_\lambda$ with variance $\sigma^2 \lambda$ to $X$. The new measurement error is $u + \tilde{u}_\lambda$ with variance $\sigma^2(1+\lambda)$, and the corresponding population slope parameter is $g(\lambda) \equiv \psi_1 \cdot \frac{var(X^*|D^*=1)}{var(X^*|D^*=1)+\sigma^2(1+\lambda)}$. By choosing different $\lambda$'s, we obtain the value of the $g$ function at various points, which we can then use to extrapolate and recover $g(-1) = \psi_1$.

---

[24] We thank a referee for suggesting this approach. See Chapter 5 of Carroll et al. (2006) for a detailed overview of the SIMEX method in the measurement error context.

The linearity of $E[Y|X^* = x^*, D^* = 1]$ in the example above can be relaxed, and Carroll et al. (1999) propose a nonparametric procedure provided that the target function is smooth. For each $\lambda$ we choose, we can estimate the value of the conditional expectation function of $\mu_d(\lambda) = E[Y|X + \tilde{u}_\lambda = 0, D^* = d]$ via a local linear regression, and use polynomials to extrapolate $\mu_d(\lambda)$ back to $\lambda = -1$ and recover the left ($d = 1$) and right ($d = 0$) intercepts in the RD estimand. The difference between $\mu_1$ and $\mu_0$ at $\lambda = -1$ identifies the RD treatment effect parameter. In this next subsection, we consider whether and how the three approaches can be extended to a design with imperfect compliance.

## 4.2 Identification under Imperfect Compliance

Only one of the three approaches proposed in section 4.1 carries over to an RD design with imperfect compliance. With the one-sided fuzzy assumption, Approach 1 still identifies the $X^*$ distribution and the RD treatment effect $\delta^c_{fuzzy}$. The idea is similar in spirit to that of Proposition 1F, which relies on Assumptions DB, 1F, 2F, 3F, 4 and 5. Analogous to Proposition 2F, we can identify the $X^*$ distribution for each value value of $D$ and $Y$ under strong independence and nondifferential measurement error, which allows the identification of $E[D|X^*]$ and $E[Y|X^*]$ by Bayes' Rule and hence the RD treatment effect.

**Proposition 4F** (a) Under Assumptions 1F, 4 and 7, the distributions of $X^*$ and $u$ are identified. (b) Under Assumptions 1FY, 2C, 4 and 7, $\delta^c_{fuzzy}$ is identified.

Given Proposition 4(a), it is straightforward to prove Proposition 4F(a). Following Schwarz and Bellegem (2010), $f_{X^*|D=1}$ and $\sigma$ are identified the same way as in Proposition 4(a). After pinning down $\sigma$, $f_{X^*|D=0}$ is identified from inverse Fourier transform $f_{X^*|D=0} = \frac{1}{2\pi} \int_{-\infty}^{\infty} e^{-itx} \frac{\varphi_{X|D=0}(t)}{\varphi_{\phi(0,\sigma^2)}(t)} dt$. The unconditional density, $f_{X^*}$ is identified by applying equation (22) with $D^*$ replaced by $D$. Finally, part (b) of Proposition 4F can be proved using equation (23) the continuous analog of (15).

To understand the role the assumptions play in Proposition 4, we make two remarks as we compare these assumptions to those of Propositions 1F and 2F. First, Assumption 4 (one-sided fuzzy) is no longer needed if the researcher knows what $\sigma$ is, which is parallel to Assumption 4's redundancy for Proposition 2F when the support of the measurement error is known. Second, the normality measurement error assumption plays a key role in Proposition 4F: It encapsulates symmetry (a stronger version of Assumption 5) and the tail restriction normality imposes renders the continuous analog of Assumption 2F unnecessary in the identification of the $X^*$ distribution.

Unfortunately, Approaches 2 and 3 do not carry over to the fuzzy case, even after assuming one-sided fuzziness. For the identification-at-infinity/boundary strategy (Approach 2), as $X$ approaches $-\infty$, the control group, $D = 0$, consists of both compliers with $X^*$ close to the threshold and never takers with a large negative $X^*$. Absent strong assumptions, it is not possible to disentangle the two groups and identify $\lim_{x^* \downarrow 0} E[Y|X^* = x^*]$. The SIMEX strategy (Approach 3) will not recover the discontinuity in the first-stage relationship $E[D|X^*]$. Without observing $D^*$, it is not possible extrapolate the left and right intercepts in the first stage relationship from separate regressions. Applying SIMEX to only the observed first-stage relationship $E[D|X]$ also will not work because the target function $E[D|X^*]$ is potentially discontinuous at zero. Despite being more nonparametric, the inability of Approaches 2 and 3 to identify the RD treatment effect in a fuzzy design greatly limits their usefulness in practice.

## 4.3 Estimation

In this section, we discuss estimation for the three approaches introduced in section 4.1. For Approach 1, Schwarz and Bellegem (2010) propose a conceptual minimum distance framework. Applied to our context, it will select the estimators for the distributions of $X^*|D^* = 1$ and $u$ to fit the observed characteristic function of $X|D^* = 1$. The fact that the distribution of $X^*$ given $D^* = 1$ has no support on the entire interval of $(0, \infty)$ satisfies the regularity condition of Theorem 3.5 of Schwarz and Bellegem (2010), which proves the consistency of the conceptual estimators. As for practical implementation, Schwarz and Bellegem (2010) suggest discretizing the $X^*|D^* = 1$ distribution with the number of support points increasing as $n \to \infty$, but defer the details to future research. Assuming that the Lapalace transform of the target density decays sufficiently quickly, Matias (2002) proposes an explicit estimator for $\sigma$ and shows that the rate of convergence is slower than $\log n$. Matias (2002) also proves that the minimax mean squared error of the pointwise density estimator cannot decrease faster than $\frac{1}{\log n}$ uniformly over a set of regular densities, which poses a challenge for most empirical applications.

For Approach 2, one can adapt the estimator proposed by Andrews and Schafgans (1998) as a starting point for estimation and inference. More precisely, let $s(\cdot)$ be a weight function that takes the value 0 for $x \leq 0$, the value 1 for $x \geq b$ for some positive constant $b$, and is smoothly increasing between 0 and $b$ with bounded derivatives. $\gamma_n$ is the bandwidth parameter, which goes to infinity with the sample size $n$. The Andrews-Schafgans estimator is defined as

$$\hat{\kappa}_1 = \frac{\sum_{i=1}^{N} Y_i D_i s(X_i - \gamma_N)}{\sum_{i=1}^{N} D_i s(X_i - \gamma_N)},$$

where $\kappa_1 \equiv \lim_{x \to \infty} E[Y|X = x, D^* = 1]$ is the quantity to be estimated. The consistency result $\hat{\kappa}_1 \xrightarrow{p} \kappa_1$ in Andrews and Schafgans (1998) is established under similar assumptions as in this paper, along with additional regularity conditions requiring 1) finite moments of the random variables and 2) an upper bound on the rate at which $\gamma_n$ tends to infinity, which depends on the upper tail probabilities of $X_i$. Furthermore, with one extra condition setting a lower bound for $\gamma_n$ as $n \to \infty$, Andrews and Schafgans (1998) also show that the distribution of the estimator $\hat{\kappa}_1$ is asymptotically normal and centered around $\kappa_1$. The rate of convergence is $\sqrt{n}$ multiplied by a factor that depends on the tail of $X$ and the exact choice of $\gamma_n$.

For Approach 3, Carroll et al. (1999) show that the error of their local linear nonparametric SIMEX estimator with bandwidth $h$ is of order $O_p\{h^2 + (nh)^{-\frac{1}{2}}\}$, provided that the polynomial extrapolant is exact. However, the performance of the estimator is sensitive in practice. Berry et al. (2002) note that alternative choices of smoothing parameters may give rise to great instability in estimates, and Staudenmayer and Ruppert (2004) find that the SIMEX estimates are not robust to alternative values within the 95% confidence interval for $\sigma$, which is estimated in an external validation study. Given the challenges in estimation described in this section, we will adopt a parametric approach in the subsequent empirical illustration.

## 5  Empirical Application: Crowdout of Private Insurance by Medicaid

As one of the largest entitlement programs in the United States, Medicaid has received considerable attention in policy discussions. An important debate is how much Medicaid eligibility crowds out private insurance–that is, the extent to which eligible individuals drop private insurance and enroll in Medicaid. If crowdout exists, the increase in the Medicaid-covered population due to an eligibility expansion will not be commensurate with the increase in overall health insurance coverage. In a seminal paper, Cutler and Gruber (1996) use simulated instruments and estimate that Medicaid reduced private insurance coverage by 30 to 40 percent for children between 1988 and 1993 when Medicaid eligibility greatly expanded. This high crowdout rate, however, is not the consensus of the literature. Several subsequent studies (e.g., Thorpe and Florence (1998), Yazici and Kaestner (2000) and Card and Shore-Sheppard (2004)) find smaller effects for the same time period using various research designs. As noted by Shore-Sheppard (2008), the question of crowdout has produced less consensus than disagreement.

In this section, we attempt to estimate Medicaid takeup and private insurance crowdout using an RD design. We use data from Card and Shore-Sheppard (2004), who study health insurance coverage and private insurance crowdout by applying an RD design around the age/birthday discontinuity for Medicaid eligibility. While not the main source of identifying variation in their study, Medicaid eligibility is also determined by an income test. If administrative income information for Medicaid applicants were available, one can conceivably apply a straightforward RD design around the relevant income thresholds to measure crowdout. In the exercise below, we explore the extent to which an RD design can be applied using noisy income measures from survey data. Specifically, we use data derived from the full panel research files of the 1990-93 SIPP and policy thresholds constructed by Card and Shore-Sheppard (2004). We make further restrictions to arrive at our analysis sample, and the details are described in Appendix E.

We apply two modeling approaches to estimate Medicaid takeup and private insurance crowdout around the income eligibility threshold. First, we use the proposed method in section 3 based on a discretized income measure. As described in section 4, nonparametric approaches (identification-at-infinity and SIMEX) cannot be applied in a fuzzy design, and the semiparametric estimator (assuming measurement error normality) has a slow convergence rate. Therefore, as our second modeling strategy, we treat income as continuous and adopt a parametric maximum likelihood estimation framework. In particular, we specify 1) $X^*$ to be the transformation of a normal random variable that allows for bunching at the income eligibility threshold, 2) $u$ to be mean zero normal, and 3) the one-sided fuzzy first stage and outcome relationships to be logistic in polynomials of $X^*$ and $D$. Details of this parametric formulation are provided in Appendix F.

As mentioned in section 3.5, we may need to transform the income variable in order for Assumptions 1FY (strong independence) and 5 (symmetry in the support of $u$) to plausibly hold. Through experimentation, we find that a Box-Cox transformation with parameter $\rho$ between 0.3 and 0.35 appears to be consistent with these assumptions as indicated by the overidentification tests. Therefore, we present the main results using $\rho = 0.33$, but also present estimates under other values of $\rho$ as robustness checks. To help understand the meaning of the transformed assignment variable, Table A.1 provides a mapping between the transformed income measures that are normalized against eligibility cutoffs and the actual family income amount. For example, a family with a child facing the 100% federal poverty line (FPL) has an actual monthly income of $1,117 (1991 dollars) if the transformed and normalized income measure is 0; the family's actual income is $1,231 if the transformed and normalized income measure takes on the value of 1. In the remainder of the section, we refer to this transformed and normalized income variable simply as income.

Figures A.1 to A.3 plot the density of the *observed* income, the first stage relationship between Medicaid coverage and income, and the analogous outcome relationship between private insurance coverage and income. The distribution of income in Figure A.1 is approximately symmetric and shows no visible sign of bunching around the cutoff. In Figure A.2, Medicaid coverage decreases as income rises, but there is no discontinuous drop-off at the cutoff, which is to be expected given the presence of measurement error. Correspondingly, Figure A.3 shows that while private insurance coverage increases with income, the relationship is again smooth through the cutoff.

Figures A.4 to A.6 plot the *estimated true* income density, and the first stage and outcome relationships. Estimates from the discrete and continuous models are juxtaposed together for ease of comparison. In A.4, while the discrete model yields a much noisier estimate of the density than the parametric continuous model, the two models roughly agree for much of the distribution and especially near the cutoff. Turning to Figure A.5, we again find that the first stage estimated using the discrete model displays a similar trend near the cutoff, albeit with more noise. The estimated first stages suggest that Medicaid coverage increases about 15 to 20 percentage points at the income threshold.[25] Finally, the behavior of the discrete model estimates in A.6 matches their continuous counterparts close to the threshold, and no visible discontinuity is detected.

The numerical discontinuity estimates are presented in Table A.2 and A.3 for the discrete and continuous models, respectively. For each combination of transformation and trimming percentage parameters (see section 3.5 for the discussion on trimming), we report in columns (1) through (4) of Table A.2 discontinuity estimates in the $X^*$ distribution, Medicaid coverage, private insurance coverage, and private insurance crowdout. Consistent with visual evidence, we generally find a statistically insignificant discontinuity in the income distribution, a significant discontinuity in Medicaid coverage between 15 and 25 percentage points, and insignificant discontinuity estimates for private insurance coverage. The resulting crowdout estimates are generally insignificant as well, but they are very imprecise–the 95% confidence intervals contain practically all estimates from the literature. In column (5), we report the p-values from the overidentification tests, and none of the models in the table are formally rejected at the 5% level.

---

[25]The one-sided fuzzy assumption is plausible in the Medicaid context. According to CMS (2014), only 3.1% of the cases were incorrectly assigned eligibility status when families applied to public health insurance, and the trimming we conduct can further alleviate the concern. We should note, however, that the ongoing Medicaid participants might not have their income eligibility recertified every month, which potentially casts doubt on the one-sided fuzzy assumption, but there is little evidence of income rebounding for them as documented by Pei (Forthcoming).

For brevity, we only present the continuous model results using the Box-Cox transformation parameter $\lambda = 0.33$ and 1% trimming in Table A.3. We find no evidence of bunching in the income distribution, a statistically significant 12.5% Medicaid takeup rate just below the eligibility cutoff, and a small and statistically insignificant crowdout effect. The parsimonious specification in the continuous model leads to much improved precision in the estimates, and even the upper bound of the 95% confidence interval excludes the crowdout estimates from Cutler and Gruber (1996).

We assess the fit of the continuous model in Table A.4 and Figure A.7. Since the model provides a parametric representation of the joint distribution $(X, D, Y)$, we examine its fit by putting the estimated parameters back to the model and gauging its success in predicting 1) the four probabilities $\Pr(D = d, Y = y)$ for $d, y = 0, 1$ and 2) the $X$ distribution within each of the four subgroups. Table A.4 compares the actual and model predicted probabilities, and the model prediction errors appear to be less than 1 percentage points for all four probabilities. Figure A.7 superimposes the model predicted $X$ distribution on top of the observed histogram of $X$ within each of the four subgroups, and in all four panels the model prediction captures the shape of the histograms.

To summarize, the discrete and continuous models provide reasonable fit to the data. However, due to the lack of functional form restrictions imposed by the discrete model, its estimates are quite noisy. Nevertheless, the two models are broadly consistent in their estimates: We find that the Medicaid takeup rate just below the eligibility cutoff falls between 10 and 25 percent and that there is little evidence supporting discontinuity in the income distribution around the cutoff and private insurance crowdout.

# 6 Conclusion

This paper investigates identification in the context of an RD design where the assignment variable is measured with error. We attempt to answer the question: what can we identify when only a noisy version of the assignment variable and the treatment status are observed? This is a challenging problem in that the presence of measurement error may smooth out the first stage discontinuity and eliminate the source of identification. Understanding this problem is important for recognizing the limitations of certain RD applications and contributes to the measurement error literature.

We first study the case where the assignment variable and the measurement error are discrete, and propose sufficient conditions to identify the assignment variable distribution using only its mismeasured counterpart and program eligibility in a sharp RD design. We then provide sufficient conditions for the identification of both the true assignment variable distribution and the first-stage and outcome relationships in the more general fuzzy RD design. A simple estimation procedure is proposed using a minimum distance framework. Following standard arguments, the resulting estimators are $\sqrt{N}$-consistent, asymptotically normal, and efficient. A numerical example verifies that the true assignment variable distribution and the RD treatment effect parameter can indeed be recovered using the proposed method.

We also explore the case where the assignment variable and measurement error are continuous and propose three different identification approaches. The first approach assumes normality of the measurement error, and identifies the assignment variable distribution and the RD treatment effect in sharp and one-sided fuzzy designs. The second approach adopts a novel identification-at-infinity strategy, and the third approach relies on the SIMEX method of Carroll et al. (1999), both of which identify the RD treatment effect in a sharp design. Because the first approach accommodates a fuzzy design, it potentially works for a larger range of applications, but its appeal is limited by the slow convergence rate of the semiparametric estimator.

In our empirical application, we apply an RD design to study Medicaid takeup and private insurance crowdout using the SIPP data from Card and Shore-Sheppard (2004). We exploit the discontinuity in the eligibility formula with respect to family income, but because income is measured with error, the first stage relationship between Medicaid takeup and reported income is not discontinuous at the eligibility cutoff. We use two approaches to recover the true income distribution and the RD treatment effect: Our proposed method from section 3 based on discretized income and, due to the difficulty in semiparametric estimation, a *parametric* MLE framework that treats income as continuous. The two approaches yield similar results: Medicaid takeup rate for the barely eligible is between 10 and 25 percent and that there is little evidence of sorting around the threshold and private insurance crowdout. However, the estimates from the discrete approach are imprecise, which by itself is unlikely to deliver convincing policy conclusions. We conclude that the parametric approach adopted in this paper for classical measurement error and in Hullegie and Klein (2010) for Berkson measurement error can be an attractive starting point for empirical researchers.

Figure 1: Theoretical Effect of Smooth Measurement Error in the Assignment Variable

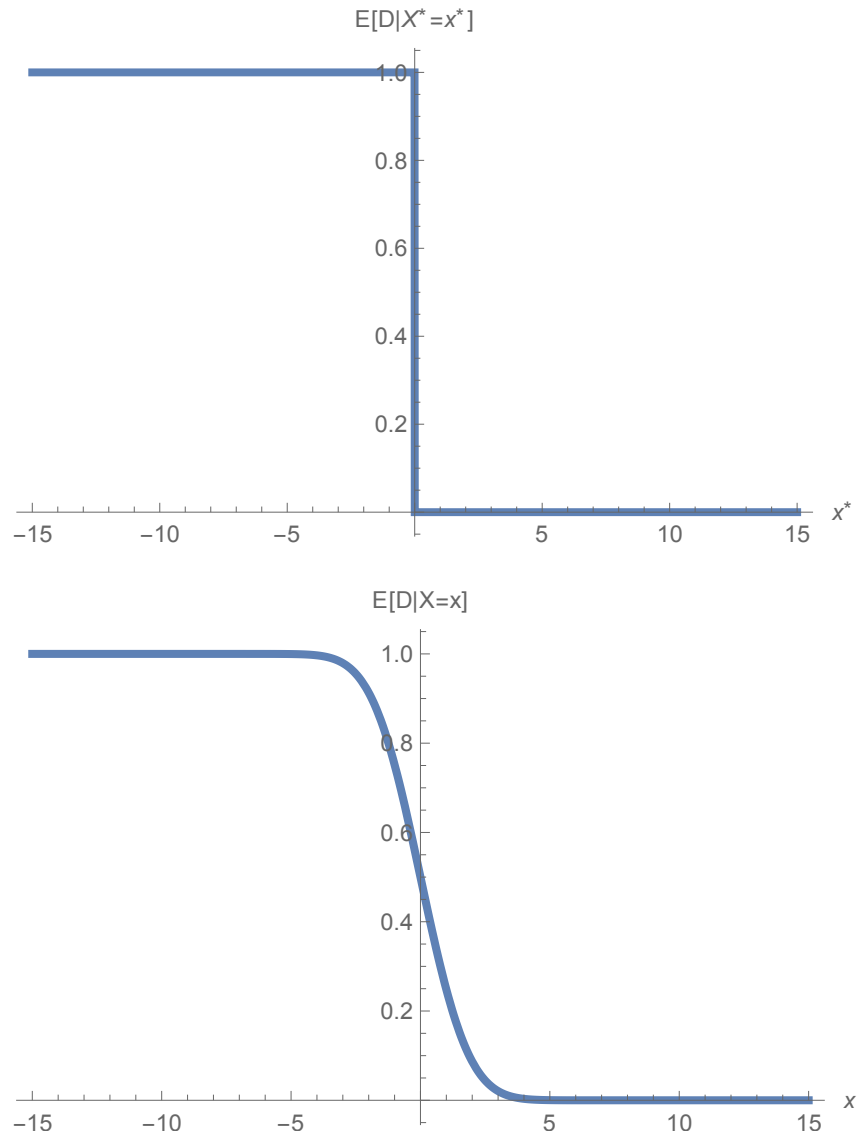

Notes: The upper panel plots the true first-stage relationship $E[D|X^* = x^*]$ in a sharp RD. The lower panel plots the observed first-stage relationship $E[D|X = x]$. The lower panel is generated by assuming that $X^*$ and $u$ are both normally distributed.

Figure 2: Observed First and Second Stage: Expectation of *D* and *Y* Conditional on the Noisy Assignment Variable *X*

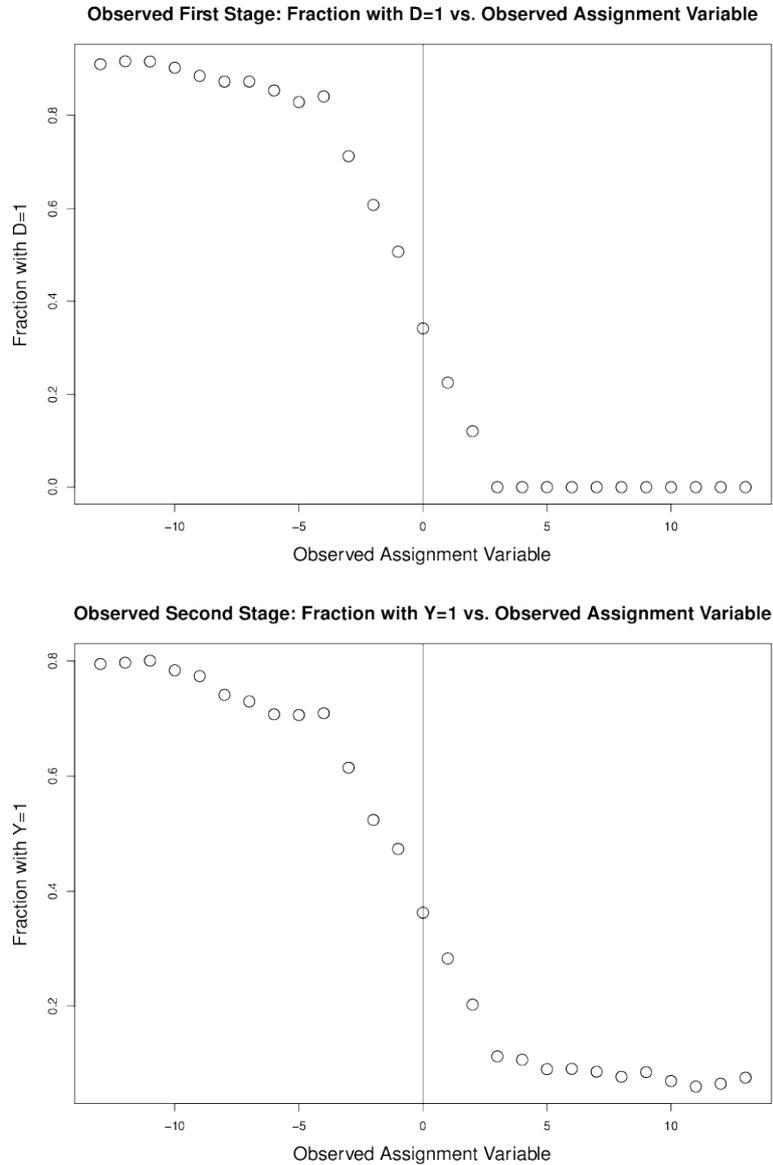

Notes: Illustrative example is based on a sample of size $N = 25,000$. $X^*$ and $u$ are uniformly distributed on the set of integers in $[-10, 10]$ and $[-3, 3]$, respectively. The true first stage and outcome response functions are $E[D|X^*] = (-0.01X^* + 0.8)D^*$ and $E[Y|X^*, D] = 0.15 + 0.6D - 0.01X^*$, respectively, which imply a true second stage equation of $E[Y|X^*] = 0.15 + 0.48D^* - 0.01X^* - 0.006D^*X^*$. Plotted are $E[D|X]$ and $E[Y|X]$ respectively where $X = X^* + u$.

Figure 3: Estimated First and Second Stage: Expectation of $D$ and $Y$ Conditional on the True Assignment Variable $X^*$

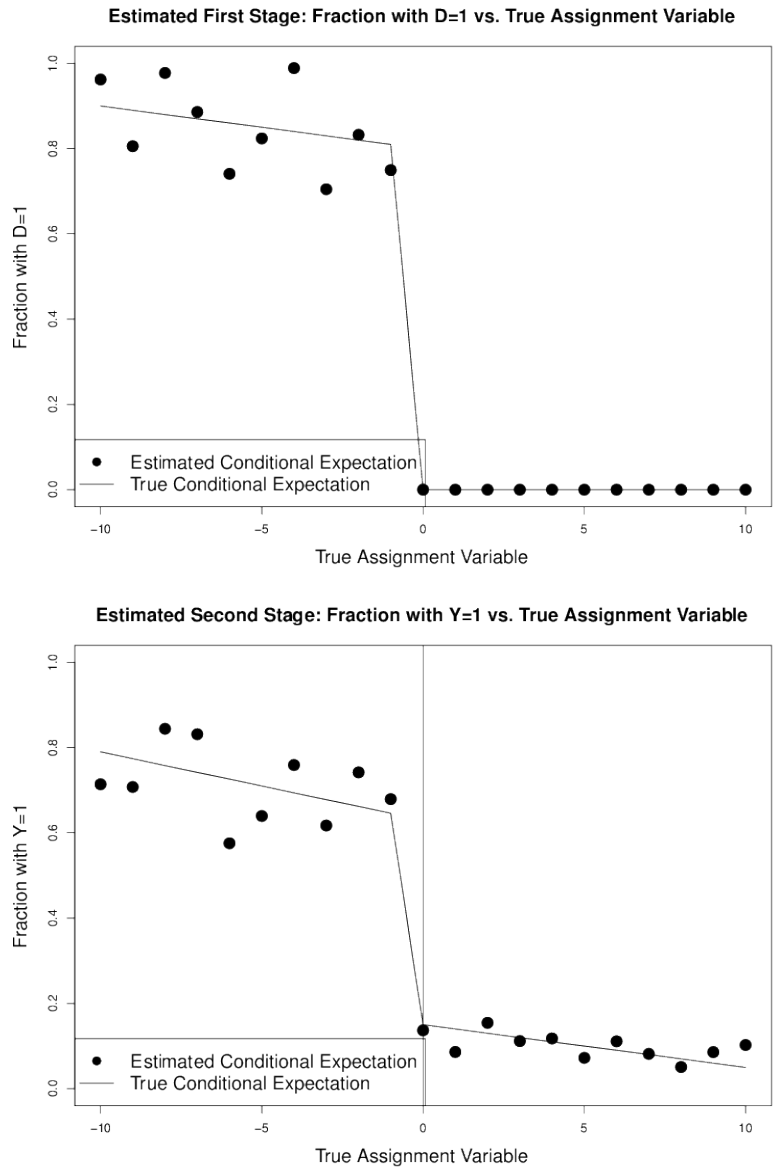

Notes: Illustrative example is based on a sample of size $N = 25,000$. $X^*$ and $u$ are uniformly distributed on the set of integers in $[-10, 10]$ and $[-3, 3]$, respectively. The true first stage and outcome response functions are $E[D|X^*] = (-0.01X^* + 0.8)D^*$ and $E[Y|X^*, D] = 0.15 + 0.6D - 0.01X^*$, respectively, which imply a true second stage equation of $E[Y|X^*] = 0.15 + 0.48D^* - 0.01X^* - 0.006D^*X^*$. Plotted are the estimated $E[Y|X^*]$ and $E[D|X^*]$ following procedures developed in section 3.4 against the true conditional expectations specified.

Figure 4: Assignment Variable Distribution with Uniform $X^*$ Distribution: Observed vs. Estimated

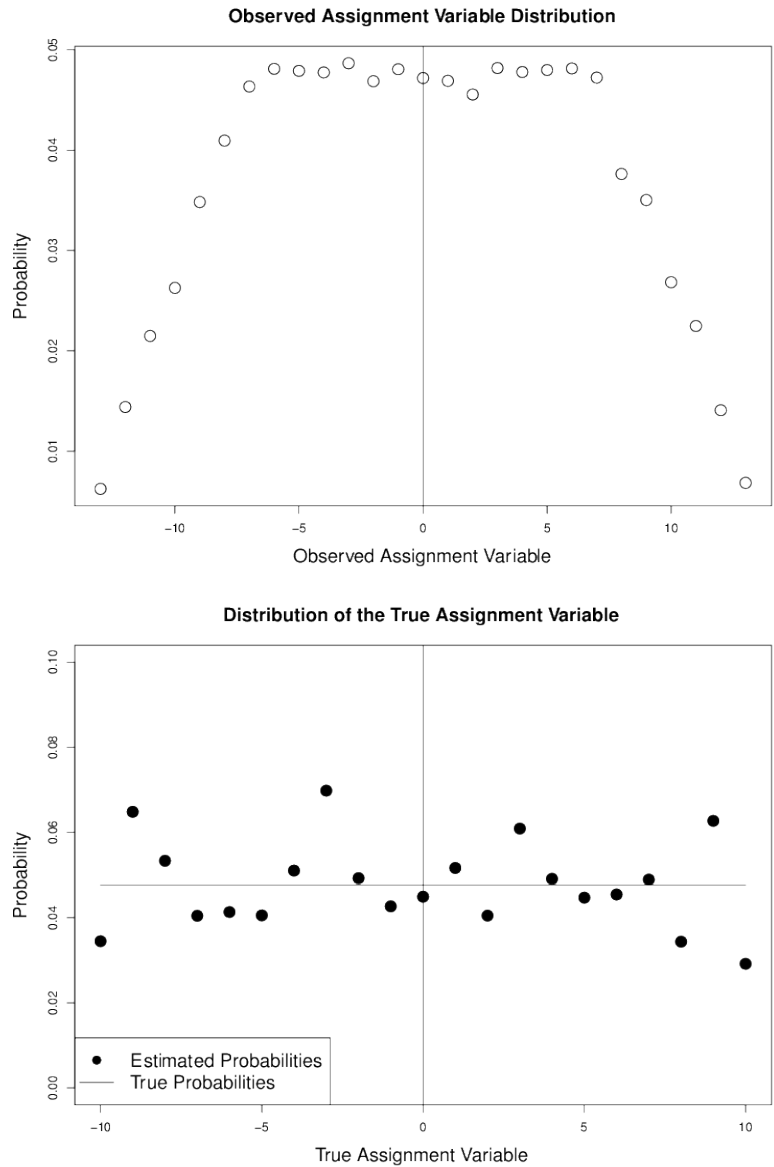

Notes: Illustrative example is based on a sample of size $N = 25,000$. $X^*$ and $u$ are uniformly distributed on the set of integers in $[-10, 10]$ and $[-3, 3]$, respectively. The true first stage and outcome response functions are $E[D|X^*] = (-0.01X^* + 0.8)D^*$ and $E[Y|X^*, D] = 0.15 + 0.6D - 0.01X^*$, respectively, which imply a true second stage equation of $E[Y|X^*] = 0.15 + 0.48D^* - 0.01X^* - 0.006D^*X^*$. Plotted are the distributions of $X$ and $X^*$, with the latter against the true uniform distribution specified.

Figure 5: Assignment Variable Distribution when True $X^*$ Distribution is Not Smooth: Observed vs. Estimated

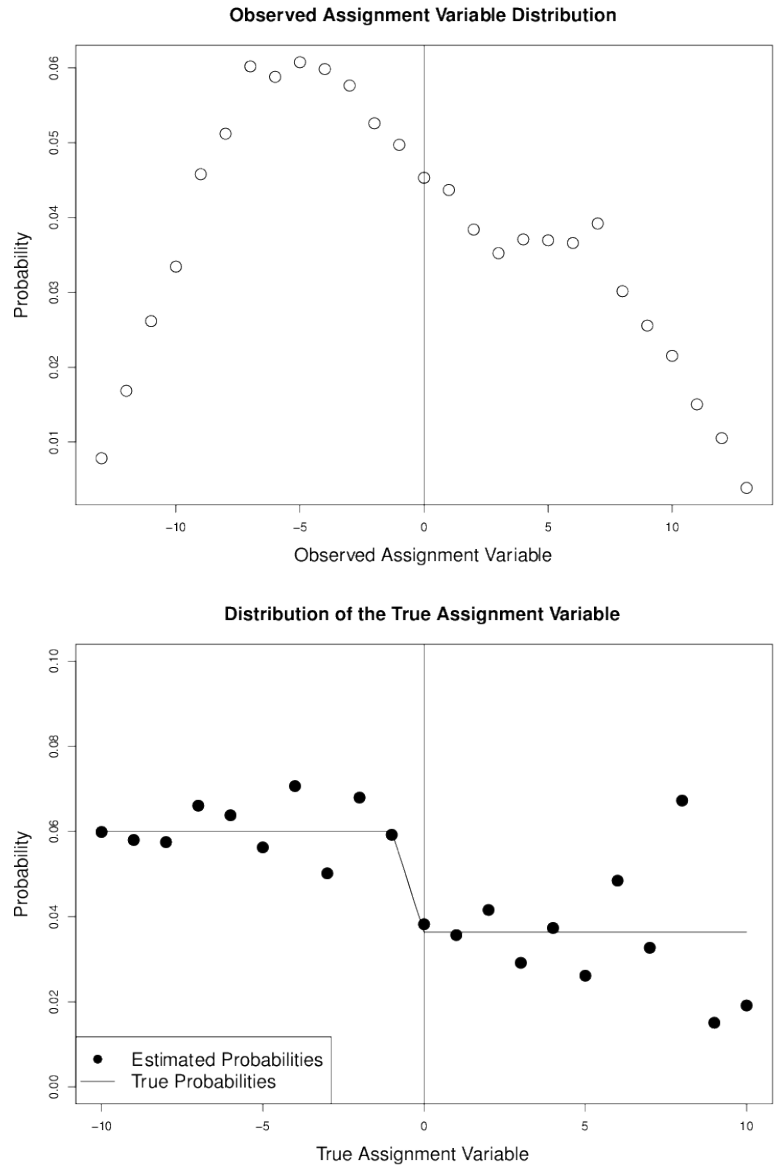

Notes: Illustrative example is based on a sample of size $N = 25,000$. $X^*$ is supported on the set of integers in $[-10, 10]$ with $\Pr(X^* = i) = 0.6$ for $i < 0$ and $\Pr(X^* = i) = 0.4$ for $i \geqslant 0$. Other specifications are the same as those underlying Figures 2, 3 and 4. Plotted are the distributions of $X$ and $X^*$, with the latter against the true distribution specified above.

Figure 6: Identification-at-infinity Illustration

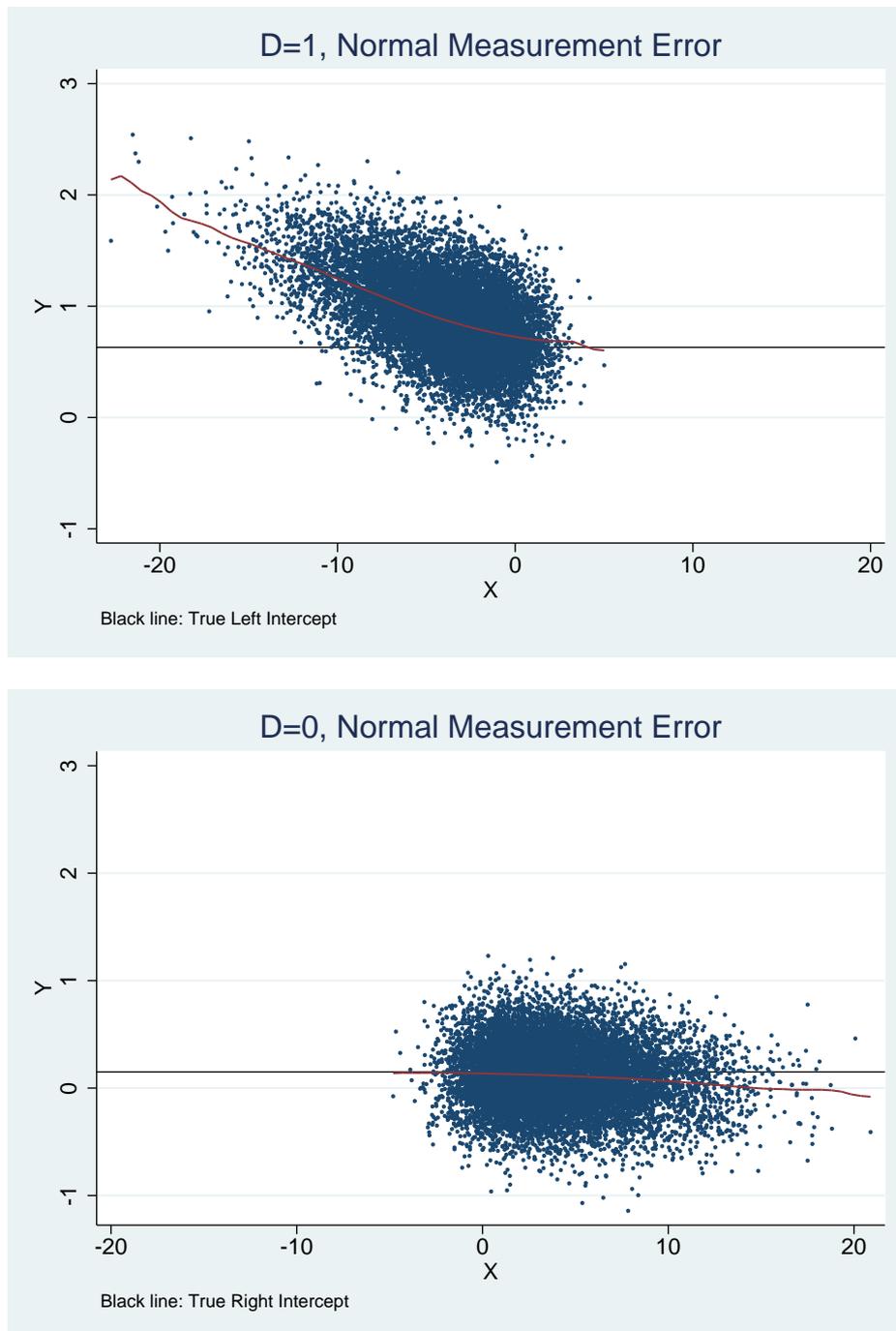

Notes: Scatter plot with 25,000 simulated observations (dots) and local constant smoothing (curve). Local smoother is implemented with the –lpoly– in Stata with default options. The data generating process is $Y = 0.15 + 0.48D^* - 0.01X^* - 0.006D^*X^* + \varepsilon$, where $X^* \sim N(0, 25)$, $\varepsilon \sim N(0, 0.09)$ and $u \sim N(0, 2)$.

Figure 7: Illustration of Identification in Yu (2012)

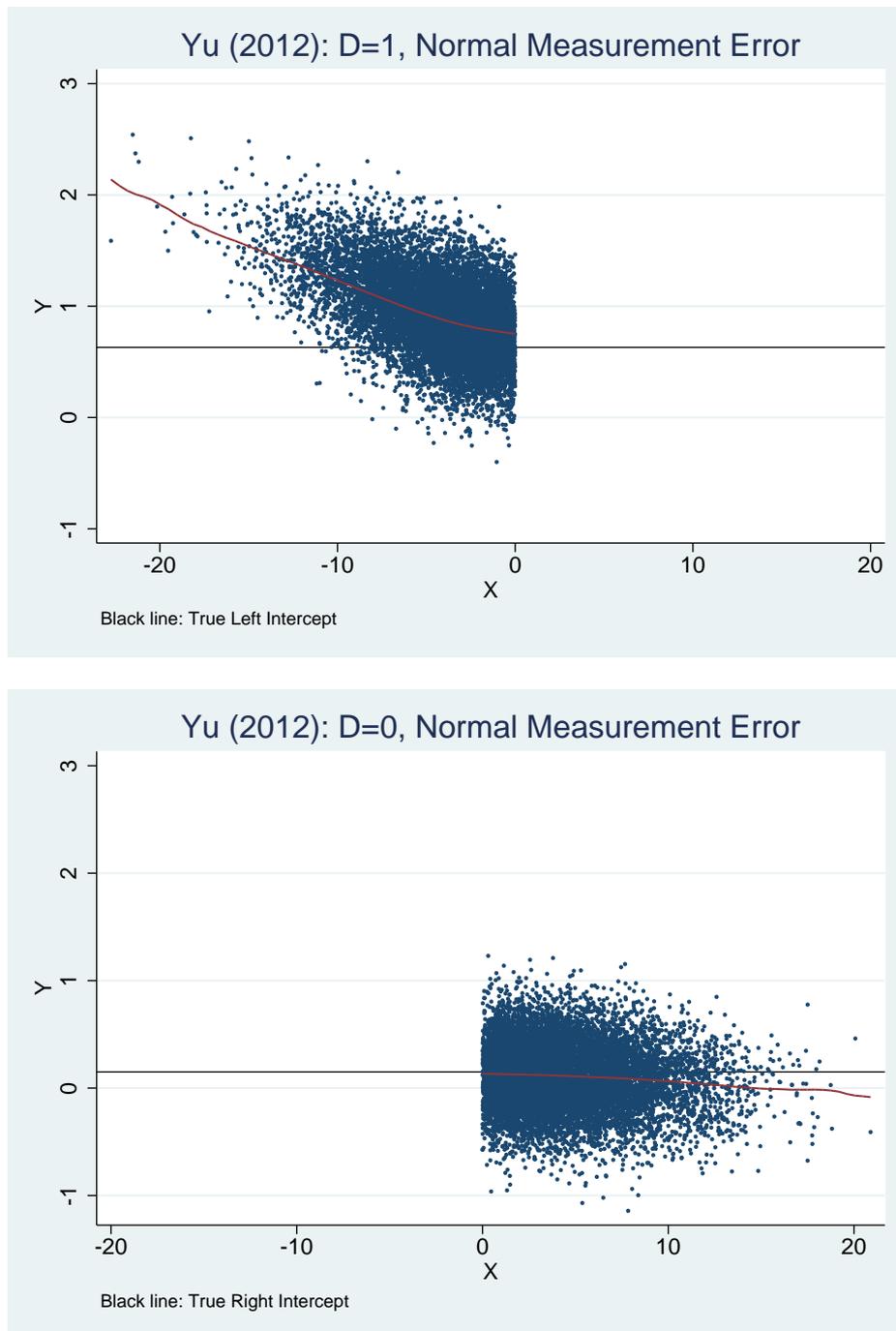

Notes: Same data and local smoothing method as Figure 6. Observations with $X \geq 0$ and $D = 1$ and with $X < 0$ and $D = 0$ are dropped per Yu (2012).

Figure 8: Illustration of Identification When the Measurement Error has Bounded Support

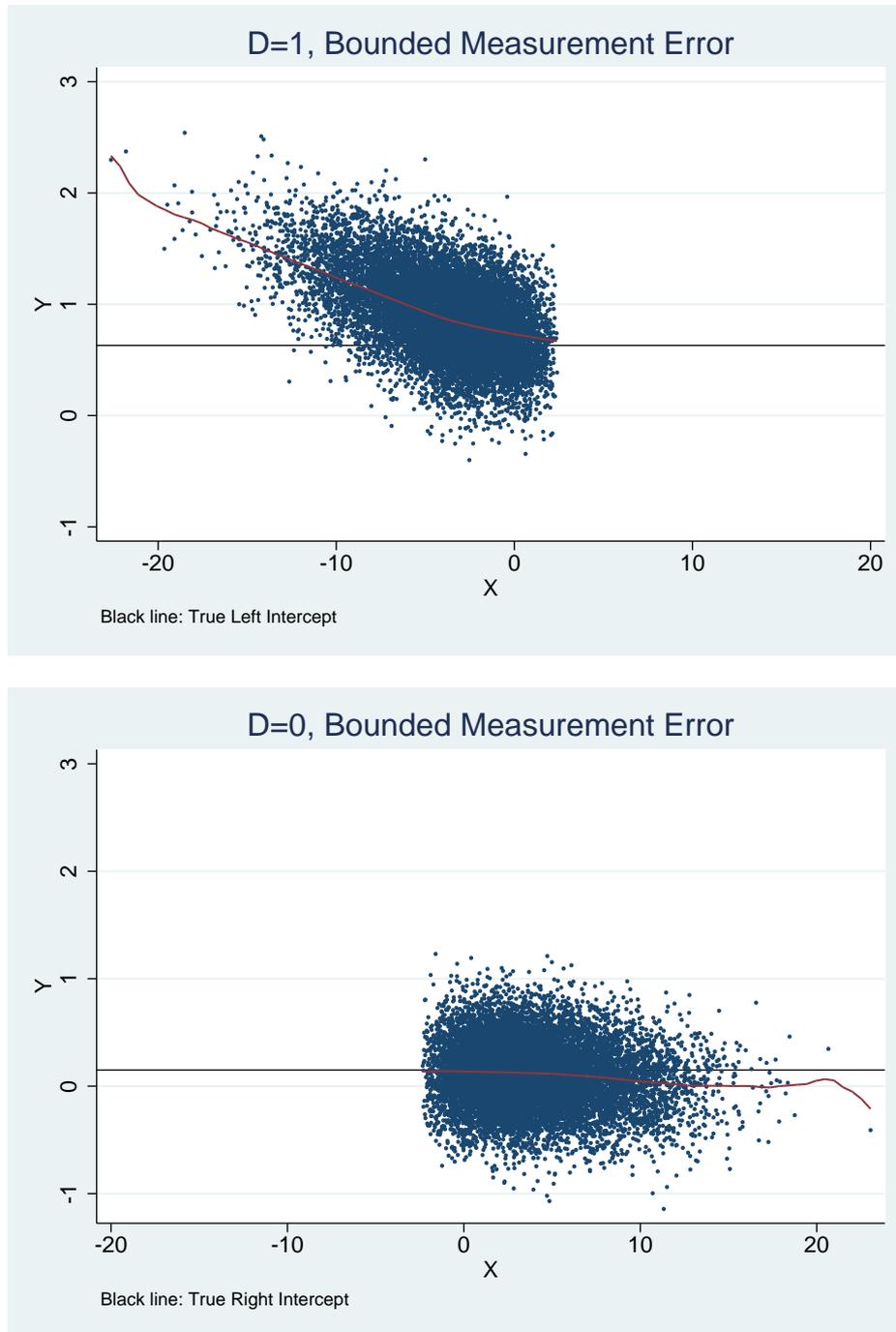

Notes: Scatter plot with 25,000 simulated observations (dots) and local constant smoothing (curve). Local smoother is implemented with the –lpoly– in Stata with default options. The data generating process is $Y = 0.15 + 0.48D^* - 0.01X^* - 0.006D^*X^* + \varepsilon$, where $X^* \sim N(0,25)$, $\varepsilon \sim N(0,0.09)$ and $u \sim unif(-\sqrt{6}, \sqrt{6})$.

# Appendix

## A  Example Documenting a Nonidentified Case in Section 3.1

Let support$_{X^*}$ = $\{-3,-2,-1,0,1,2\}$, the vectors of probability masses $(p^1_{-3}, p^1_{-2}, p^1_{-1}) = (p^0_0, p^0_1, p^0_2) = (\frac{1}{4}, \frac{1}{4}, \frac{1}{2})$ and $r^1 = \frac{1}{2}$. Let support$_u = \{-1, 0, 1\}$; and $(m_{-1}, m_0, m_1) = (\frac{1}{2}, \frac{1}{4}, \frac{1}{4})$. It follows that the observed vectors of probabilities are $(q^1_{-4}, q^1_{-3}, q^1_{-2}, q^1_{-1}, q^1_0) = (q^0_{-1}, q^0_0, q^0_1, q^0_2, q^0_3) = (\frac{1}{8}, \frac{3}{16}, \frac{3}{8}, \frac{3}{16}, \frac{1}{8})$, and the resulting $9 \times 6$ matrix

$$\mathbf{Q} = \begin{bmatrix} \frac{1}{8} & 0 & 0 & -\frac{1}{8} & 0 & 0 \\ \frac{3}{16} & \frac{1}{8} & 0 & -\frac{3}{16} & -\frac{1}{8} & 0 \\ \frac{3}{8} & \frac{3}{16} & \frac{1}{8} & -\frac{3}{8} & -\frac{3}{16} & -\frac{1}{8} \\ \frac{3}{16} & \frac{3}{8} & \frac{3}{16} & -\frac{3}{16} & -\frac{3}{8} & -\frac{3}{16} \\ \frac{1}{8} & \frac{3}{16} & \frac{3}{8} & -\frac{1}{8} & -\frac{3}{16} & -\frac{3}{8} \\ 0 & \frac{1}{8} & \frac{3}{16} & 0 & -\frac{1}{8} & -\frac{3}{16} \\ 0 & 0 & \frac{1}{8} & 0 & 0 & -\frac{1}{8} \\ 1 & 1 & 1 & 0 & 0 & 0 \\ 0 & 0 & 0 & 1 & 1 & 1 \end{bmatrix}$$

is only of rank 4. The result of nonidentification is intuitive because we can "switch" the $p$ and $m$ vectors and the alternative distributions $(\tilde{p}^1_{-3}, \tilde{p}^1_{-2}, \tilde{p}^1_{-1}) = (\tilde{p}^0_0, \tilde{p}^0_1, \tilde{p}^0_2) = (\frac{1}{2}, \frac{1}{4}, \frac{1}{4})$ and $(\tilde{m}_{-1}, \tilde{m}_0, \tilde{m}_1) = (\frac{1}{4}, \frac{1}{4}, \frac{1}{2})$ give rise to the same distributions of $X|D^* = 1$ and $X|D^* = 0$ as $(p^1_{-3}, p^1_{-2}, p^1_{-1})$, $(p^0_0, p^0_1, p^0_2)$ and $(m_{-1}, m_0, m_1)$.

# B  Identification of the Measurement Error Distribution in Lemma 2

The $m_l$'s are identified after the $p_k^1$'s and the $p_k^0$'s are identified because they solve the following linear system:

$$
\begin{bmatrix}
p_{U_{X^*}}^1 & 0 & \cdots & 0 \\
p_{U_{X^*}-1}^1 & p_{U_{X^*}}^1 & \cdots & 0 \\
\vdots & p_{U_{X^*}-1}^1 & \cdots & \vdots \\
p_0^1 & \vdots & \cdots & 0 \\
0 & p_0^1 & \cdots & p_{U_{X^*}}^1 \\
\vdots & 0 & \cdots & p_{U_{X^*}-1}^1 \\
\vdots & \vdots & \cdots & \vdots \\
0 & 0 & \cdots & p_0^1 \\
p_{-1}^0 & 0 & \cdots & 0 \\
p_{-2}^0 & p_{-1}^0 & \cdots & \vdots \\
\vdots & p_{-2}^0 & \cdots & \vdots \\
p_{L_{X^*}}^0 & \vdots & \cdots & 0 \\
0 & p_{L_{X^*}}^0 & \cdots & p_{-1}^0 \\
\vdots & 0 & \cdots & p_{-2}^0 \\
\vdots & \vdots & \cdots & \vdots \\
0 & 0 & \cdots & p_{L_{X^*}}^0
\end{bmatrix}
\underbrace{\begin{bmatrix} m_{U_u} \\ m_{U_u-1} \\ \vdots \\ m_0 \\ \vdots \\ m_{L_u+1} \\ m_{L_u} \end{bmatrix}}_{K_u \times 1}
=
\underbrace{\begin{bmatrix} q_{U_{X^*}+U_u}^1 \\ q_{U_{X^*}+U_u-1}^1 \\ \vdots \\ q_{L_u}^1 \\ q_{U_u-1}^0 \\ q_{U_u-2}^0 \\ \vdots \\ q_{L_u+L_{X^*}}^0 \end{bmatrix}}_{(K_{X^*}+2K_u-2)\times 1}
\quad (29)
$$

$$\underbrace{\phantom{X}}_{(K_{X^*}+2K_u-2)\times K_u}$$

Denote system (29) with the compact notation $\mathbf{Pm} = \mathbf{q}$, where $\mathbf{P}$ is the $(K_{X^*} + 2K_u - 2) \times K_u$ matrix containing the already known $p_k^1$ and $p_k^0$'s, $\mathbf{m}$ is the $K_u \times 1$ vector containing the $m_l$'s, and $\mathbf{q}$ is the $(K_{X^*} + 2K_u - 2) \times 1$ vector containing the constant $q_i^1$'s and $q_i^0$'s. The fact $r^1, r^0 > 0$ implies that $K_{X^*} \geq 2$, and $K_u \geq 1$ by construction. Together, they imply that $K_{X^*} + 2K_u - 2 > K_u$, which means that there are more rows than columns in $\mathbf{P}$. Because $P_k^1 > 0$ for some $k$, the columns in $\mathbf{P}$ are linearly independent. Therefore, any solution that solves (29) is unique, and the parameters $m_l$'s are consequently identified by solving (29).

# C  Assignment Variable and Measurement Error Have Discrete and Unbounded Support

Following the identification result in section 3.1, a natural question arises: how does the result extend to the case where the support of the discrete assignment variable is unbounded? While a sufficient condition for identification is left for future research, we show in this section that the model is not always identified by constructing two sets of observationally equivalent distributions. The general nonidentifiability result may not be surprising given the absence of an infinite-support-counterpart to Assumption 3, but the construction of the example is not straightforward as the technique used in the construction of a not-full-rank **Q** in the Appendix no longer applies when supports of $X^*$ and $u$ are infinite.

We construct two sets of infinitely supported distributions of $X^*$ and $u$ that are observationally equivalent, i.e. they give rise to the same joint distribution $(X, D^*)$. In particular, we specify discrete probability mass functions, $\{p_a^1, p_a^0, m_a\}$ and $\{p_b^1, p_b^0, m_b\}$ (where $p_j^1$ and $p_j^0$ ($j = a, b$) denote the conditional probability mass functions of $X^*|D^* = 1$ and $X^*|D^* = 0$ respectively) such that

1. the support of $p_a^1, p_b^1$ is the set of negative integers $\{-1, -2, -3, ...\}$;

2. the support of $p_a^0, p_b^0$ is the set of nonnegative integers $\{0, 1, 2...\}$;

3. $q^1 \equiv p_a^1 * m_a = p_b^1 * m_b$ and $q^0 \equiv p_a^0 * m_a = p_b^0 * m_b$ where $*$ denotes convolution.

As in the previous section, the probability mass functions $q^1$ and $q^0$ are the observed distribution of the noisy assignment variable $X$ conditional on $D = 1$ and $D = 0$ respectively. Note that Assumptions 1 and 2 still hold in the construction of the example.

It is useful to consider yet again the moment generating functions of the distributions, which we denote by $\{P_a^1(t), P_a^0(t), M_a(t)\}$ and $\{P_b^1(t), P_b^0(t), M_b(t)\}$.[26] Again, we can translate the convolutions of the distributions $p_a^1 * m_a = p_b^1 * m_b$ and $p_a^0 * m_a = p_b^0 * m_b$ into products of MGF's $P_a^1(t)M_a(t) = P_b^1(t)M_b(t)$ and

---

[26]When the support is unbounded, the question of convergence naturally arises regarding the moment generating functions. We are not concerned with the convergence issue and use the MGF's in the formal sense as we are only interested in the coefficients of the $e^{ti}$ terms.

$P_a^0(t)M_a(t) = P_b^0(t)M_b(t)$. It follows then that

$$P_b^1(t) = P_a^1(t)\frac{M_a(t)}{M_b(t)} \quad (30)$$
$$P_b^0(t) = P_a^0(t)\frac{M_a(t)}{M_b(t)}$$

Loosely speaking, the supports of $p_a^1$ and $p_b^1$ are preserved under convolution with the "distribution" represented by $\frac{M_a}{M_b}(t)$.

To construct the two sets of distributions, we first specify $\frac{M_a}{M_b}(t)$, $P_a^1(t)$, $P_a^0(t)$ and $M_b(t)$, and then show that $P_b^1(t)$ and $P_b^0(t)$ obtained following (30) are moment generating functions for valid probability distributions that are supported on the negative and nonnegative integers respectively. Finally, we check that $M_a(t)$ constructed by $\frac{M_a}{M_b}(t)M_b(t)$ represents a valid probability distribution.

Let

$$\frac{M_a}{M_b}(t) = c_{a/b}(x + \sum_{n\neq 0}(-x)^{|n|-1}e^{tn})$$
$$P_a^1(t) = c_a^1(\frac{x^2}{1+x^2}e^{-t} + \sum_{n\leqslant -2} x^{|n|-1}e^{tn})$$
$$P_a^0(t) = c_a^0(\frac{x^2}{1+x^2} + \sum_{n\geqslant 1} x^{|n|}e^{tn})$$
$$M_b(t) = \frac{1}{2}(P_a^1(t) + P_a^0(t))$$

where $x$ is any *constant* in the interval $(0,1)$, and $c_{a/b} = \frac{x+1}{x^2+x+2}$, $c_a^1 = c_a^0 = \frac{1-x+x^2-x^3}{x+x^2}$ are normalizing constants so that $\frac{M_a}{M_b}(0) = P_a^1(0) = P_a^0(0) = 1$ (and consequently $M_b(0) = 1$). Using (30), we obtain

$$P_b^1(t) = c_{a/b}c_a^1\left(xe^{-t} + \sum_{n\leqslant -2 \text{ and } n \text{ even}} \frac{x^{|n|}(x^2+3)}{x^2+1}e^{tn} + \sum_{n\leqslant -3 \text{ and } n \text{ odd}} (x^{|n|}+x^{|n|-2})e^{tn}\right)$$
$$P_b^0(t) = c_{a/b}c_a^0\left(x + \sum_{n\geqslant 1 \text{ and } n \text{ odd}} \frac{x^{|n|+1}(x^2+3)}{x^2+1}e^{tn} + \sum_{n\geqslant 2 \text{ and } n \text{ even}} (x^{|n|+1}+x^{|n|-1})e^{tn}\right)$$
$$M_a(t) = \frac{1}{2}[P_b^1(t) + P_b^0(t)]$$

Note that $P_b^1(t)$ only contains negative powers of $e^t$ and that $P_b^0(t)$ only contains nonnegative powers of $e^t$. Also, all coefficients of powers of $e^t$ in $P_b^1$, $P_b^0$ and $M_a$ are strictly positive with $P_b^1(0) = P_b^0(0) = M_a(0) = 1$.

Thus, $P_b^1$, $P_b^0$ and $M_a$ represent valid probability distributions that satisfy the support requirement mentioned above. Hence, (2) is not always identified when the supports of $X^*$ and $u$ are infinite.

## D  Proof of Proposition 5

In this section, we supply the proof to Proposition 5. We will prove the following lemma, which Proposition 5 trivially follows.

**Lemma 3** Suppose Assumption 2C holds and that $f_{X^*}$ is continuous at 0. If $F_u$ satisfies

$$\lim_{x \to \infty} \frac{1 - F_u(x+v)}{1 - F_u(x)} = 0 \qquad \text{for all } v > 0,$$

then

$$\lim_{x \to \infty} E[Y|X=x, D^*=1] = \lim_{x^* \uparrow 0} E[Y|X^*=x^*]. \tag{31}$$

Symmetrically, if

$$\lim_{x \to -\infty} \frac{F_u(x-v)}{F_u(x)} = 0 \qquad \text{for all } v > 0,$$

then

$$\lim_{x \to -\infty} E[Y|X=x, D^*=0] = \lim_{x^* \downarrow 0} E[Y|X^*=x^*]. \tag{32}$$

*Proof.* By symmetry, it suffices to prove (31). To this end, note that

$$E[Y|X=x, D^*=1]$$
$$=E[Y|X^*+u=x, X^*<0]$$
$$=E[E[Y|X^*,u]|X^*+u=x, X^*<0]$$
$$=E[g(X^*)|X^*+u=x, X^*<0],$$

where $g(x^*) \equiv E[Y|X^*=x^*] = E[Y|X^*=x^*, u]$ since $Y$ and $u$ are independent. The conditional density of $X^*$, given $X^*+u=x$ and $X^*<0$, is

$$\varphi(-v) = \frac{f_{X^*}(-v)f_u(x+v)}{\int_0^\infty f_{X^*}(-v)f_u(x+v)dv}, \quad v > 0.$$

Therefore

$$E[g(X^*)|X^* + u = x, X^* < 0]$$
$$= \int_0^\infty g(-v)\varphi(-v)dv$$
$$= \int_0^\infty g(-v)d\mu_x,$$

where $\mu_x$ is the conditional distribution of $-X^*$ given $X^* + u = x$ and $X^* < 0$, with density function $\varphi$.

Denote by $\eta_x$ the conditional distribution of $u - x$ given $u - x > 0$, then it has density $\psi(v) = \frac{f_u(x+v)}{1-F_u(x)}$, $v > 0$. Thus $\mu_x$ is absolutely continuous with respect to $\eta_x$, with the density

$$\frac{d\mu_x}{d\eta_x}(v) = \frac{\varphi(-v)}{\psi(v)} = \frac{f_{X^*}(-v)}{\int_0^\infty f_{X^*}(-v)\frac{f_u(x+v)}{1-F_u(x)}dv}.$$

Since $f_{X^*}$ is continuous at 0 and $f_{X^*}(0) > 0$, there exist $\varepsilon > 0, \delta > 0$, such that $f_{X^*}(-v) \geq \varepsilon$ for all $v \in [0, \delta]$.
Therefore

$$\int_0^\infty f_{X^*}(-v)\frac{f_u(x+v)}{1-F_u(x)}dv$$
$$\geq \int_0^\delta \varepsilon \frac{f_u(x+v)}{1-F_u(x)}dv$$
$$= \varepsilon \frac{F_u(x+\delta) - F_u(x)}{1 - F_u(x)}.$$

By assumption,
$$\frac{F_u(x+\delta) - F_u(x)}{1 - F_u(x)} \to 1 \quad \text{as } x \to \infty.$$

Thus for $x$ large enough, $\frac{F_u(x+\delta)-F_u(x)}{1-F_u(x)}$ is bounded from below by a positive constant. Hence

$$\frac{d\mu_x}{d\eta_x}(v) \leq c \cdot f_{X^*}(-v)$$

for some constant $c$ when $x$ is large enough. For any $a > 0$,

$$\eta_x([0,a]) = \int_0^a \psi(v)dv = \frac{F_u(x+a) - F_u(x)}{1 - F_u(x)} = 1 - \frac{1 - F_u(x+a)}{1 - F_u(x)} \to 1 \quad \text{as } x \to \infty$$

by assumption. Hence $\eta_x$ converges in distribution to Dirac measure $\delta_0$ as $x \to \infty$. Absolute continuity of $\mu_x$

with respect to $\eta_x$ then implies that $\mu_x$ converge in distribution to $\delta_0$ as well. As a result,

$$\lim_{x \to \infty} E[Y|X = x, D^* = 1]$$
$$= \lim_{x \to \infty} E[g(X^*)|X^* + u = x, X^* < 0]$$
$$= \lim_{x \to \infty} \int_0^\infty g(-v) d\mu_x$$
$$= \lim_{x^* \uparrow 0} g(x^*)$$
$$= \lim_{x^* \uparrow 0} E[Y|X^* = x^*].$$

□

# E  Analysis Sample Construction for Empirical Illustration

Moreover, the interaction between AFDC and Medicaid obscures the causal effect of taking up Medicaid on private insurance coverage. For those whose Medicaid income eligibility threshold coincides with that of the AFDC, any difference in private insurance coverage for groups just above and below the threshold is attributed to the combined receipt of Medicaid and AFDC. Therefore, it will be theoretically appealing to focus on those whose Medicaid threshold is higher than their families' AFDC cutoff for the gross income test, because they will be ineligible for the AFDC benefit if their family's income is right below the Medicaid threshold. This restriction reduces the sample size from 55,021 to 12,534, and the lowest Medicaid threshold in this sample is 100% of the federal poverty line.

I further restrict the sample by dropping the children for whom the reported family income is zero. By matching SIPP to Social Security Summary Earnings Records, Pedace and Bates (2000) find that 89% of the SIPP respondents who report zero earnings had zero earnings. Therefore, one may suspect the measurement error for those who report zero earnings will be drastically different from those who do not. Making this restriction reduces the sample size to 11,376.

# F   Formulation of the Parametric Model in Section 5

In our parametric framework that treats $X^*$ (and $u$) as continuous, $X^*$ is specified as a transformation of a smoothly distributed random variable $S$:

$$X^* = \begin{cases} S & \text{if } S \leqslant 0 \\ 0 & \text{with prob p if } S > 0 \\ S & \text{with prob(1-p) if } S > 0 \end{cases} \quad . \tag{33}$$

This specification allows for bunching in $X^*$ at the income threshold, which is motivated by neoclassical labor supply models (e.g. Saez (2010), Kleven and Waseem (2013), and see Jales and Yu (2016) for a review). The parameter $p$, which potentially depends on the value of $S$, denotes the degree of bunching–there is no bunching or discontinuity when $p = 0$. For simplicity, we assume that $S$ is normally distributed with mean $\mu$ and variance $\sigma_S^2$, but we can be more flexible by assuming a mixture normal distribution that is less restrictive. Together, with equation (33) and the normality assumption for $S$, we can express the $F_{X^*}$ using $p$, $\mu$, $\sigma_S$. For ease of exposition, we will use $f_{X^*}$ to denote the corresponding "density" of $X^*$, which invokes the Dirac Delta function to account for bunching at zero.

To be consistent with the discrete formulation, we assume a one-sided fuzzy design and impose a logit functional form in the first-stage and outcome relationships:

$$\Pr(D = 1|X^*) = \frac{1}{1+e^{-\Sigma \alpha_k X^{*k}}} 1_{[X^*<0]}$$

$$\Pr(Y = 1|X^*) = \frac{1}{1+e^{-\Sigma \beta_k X^{*k} + \delta D + \Sigma \gamma_k D \cdot X^{*k}}}.$$

Maintaining the classical measurement error assumption, we can write down the likelihood for each observation $(X_i, D_i, Y_i)$

$$\begin{aligned} L(X_i, D_i, Y_i) &= \prod_{d,y} [f_{X|D,Y}(X_i|D_i = d, Y_i = y) \Pr(D_i = d, Y_i = y)]^{1_{[D_i=d, Y_i=y]}} \\ &= \prod_{d,y} [\int f_{X^*|D,Y}(x^*|D_i = d, Y_i = y) f_u(X_i - x^*) dx^* \Pr(D_i = d, Y_i = y)]^{1_{[D_i=d, Y_i=y]}} \end{aligned} \tag{34}$$

where

$$f_{X^*|D,Y}(x^*|D_i = d, Y_i = y) = \frac{\Pr(Y_i = y|X^* = x^*, D_i = d) \Pr(D_i = d|X^* = x^*) f_{X^*}(x^*)}{\int \Pr(Y_i = y|X^* = x^*, D_i = d) \Pr(D_i = d|X^* = x^*) f_{X^*}(x^*) dx^*}$$

Figure A.1: Distribution of Observed Income

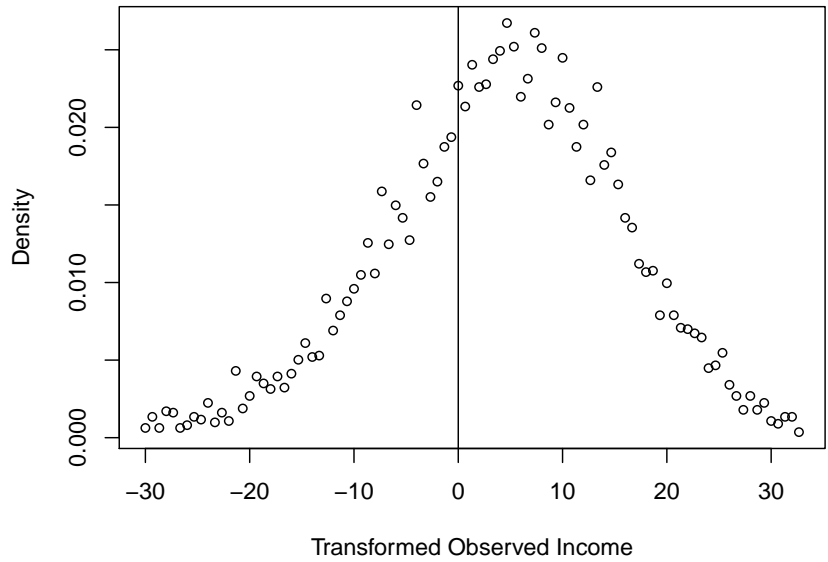

Note: Income is derived from a Box-Cox transformation of the actual family income with parameter 0.33, and is normalized against the transformed Medicaid Eligibility Threshold. Top and Bottom 1% of the normalized income is trimmed. Zero is the cutoff and a child is Medicaid eligible if the normalized income is less than zero. See Table A.1 for a mapping between the transformed and actual income values for various Medicaid cutoffs.

Figure A.2: Fraction on Medicaid vs. Observed Income

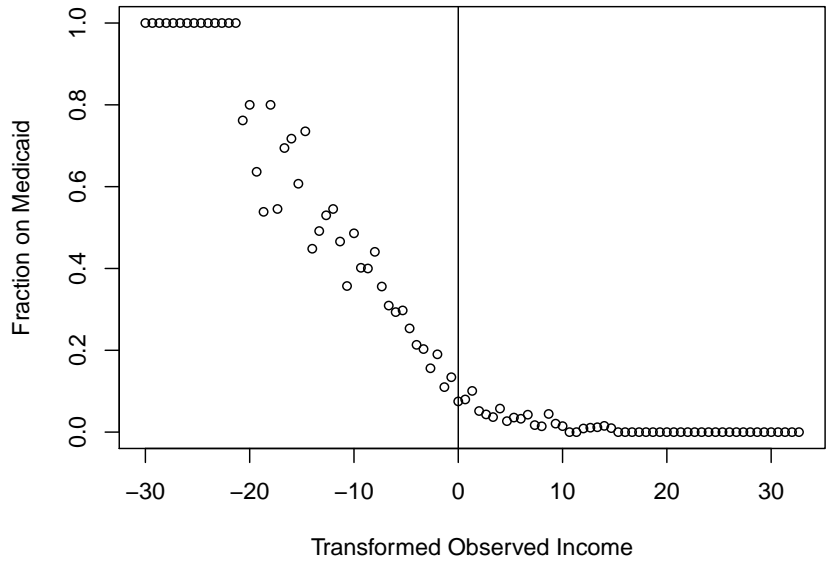

Note: Income is derived from a Box-Cox transformation of the actual family income with parameter 0.33, and is normalized against the transformed Medicaid Eligibility Threshold. Top and Bottom 1% of the normalized income is trimmed. Zero is the cutoff and a child is Medicaid eligible if the normalized income is less than zero. See Table A.1 for a mapping between the transformed and actual income values for various Medicaid cutoffs.

Figure A.3: Fraction on Private Insurance vs. Observed Income

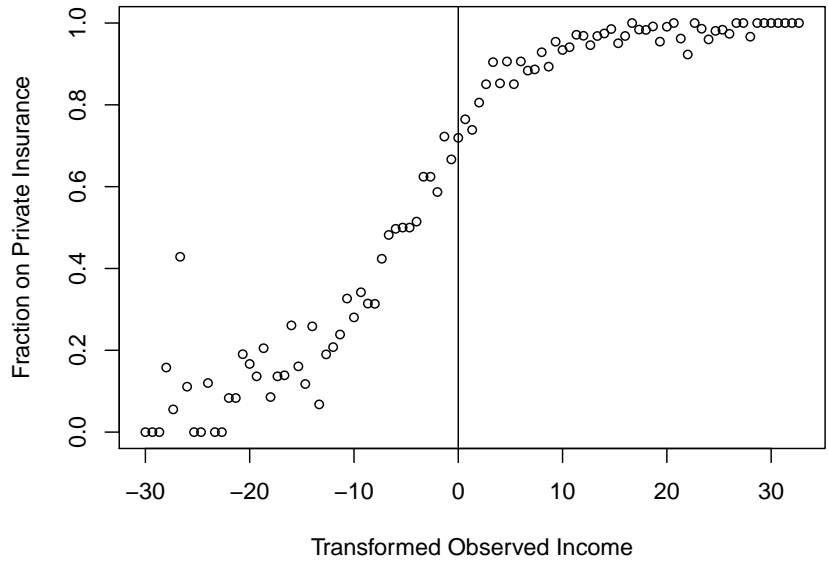

Note: Income is derived from a Box-Cox transformation of the actual family income with parameter 0.33, and is normalized against the transformed Medicaid Eligibility Threshold. Top and Bottom 1% of the normalized income is trimmed. Zero is the cutoff and a child is Medicaid eligible if the normalized income is less than zero. See Table A.1 for a mapping between the transformed and actual income measures for various Medicaid cutoffs.

Figure A.4: Distribution of True Income

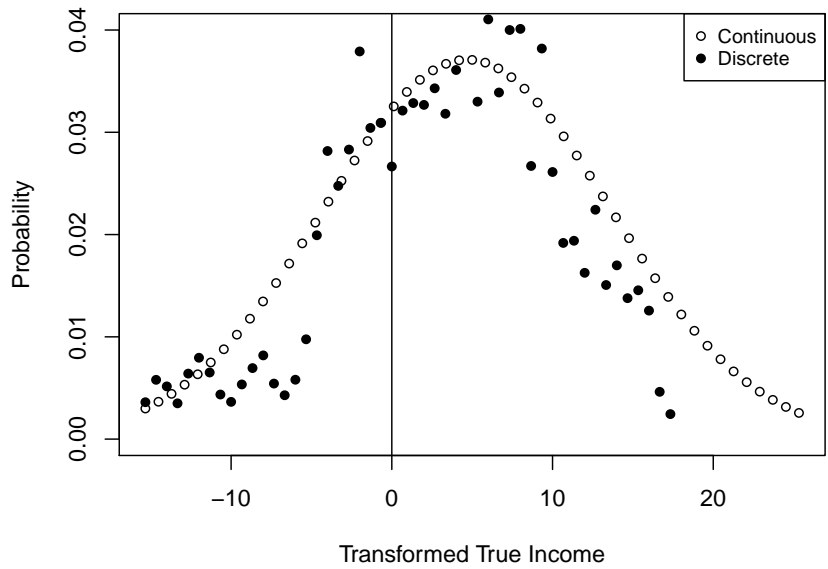

Note: This figure plots the estimated distribution of transformed true income by following the discrete (solid points) and continuous (hollow points) modeling approaches described in sections 3 and 5. See Table A.1 for a mapping between the transformed and actual income measures for various Medicaid cutoffs.

Figure A.5: Fraction on Medicaid vs. True Income

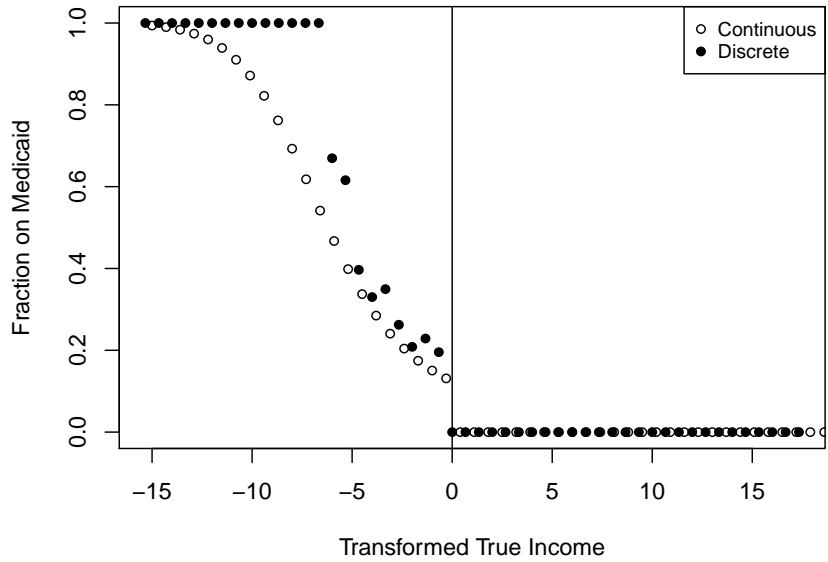

Note: This figure plots the estimated distribution of transformed true income by following the discrete (solid points) and continuous (hollow points) modeling approaches described in sections 3 and 5. See Table A.1 for a mapping between the transformed and actual income measures for various Medicaid cutoffs.

Figure A.6: Fraction on Private Insurance vs. True Income

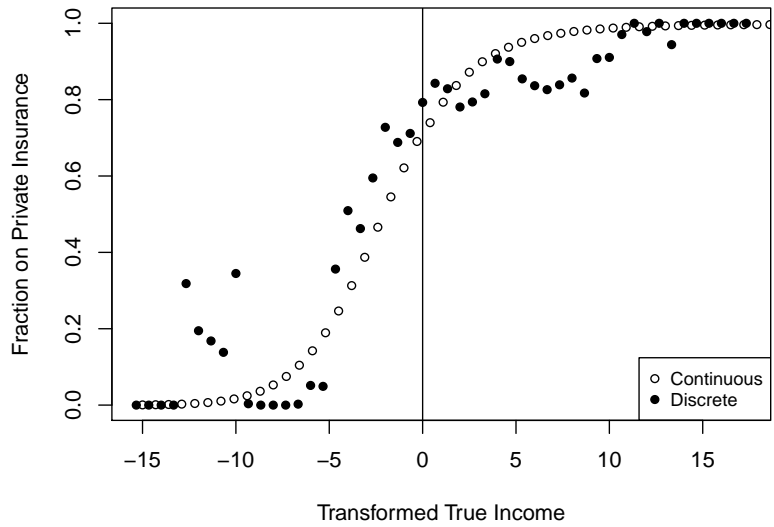

Note: This figure plots the estimated distribution of transformed true income by following the discrete (solid points) and continuous (hollow points) modeling approaches described in sections 3 and 5. See Table A.1 for a mapping between the transformed and actual income measures for various Medicaid cutoffs.

Figure A.7: Actual vs. Model Predicted Income Distribution

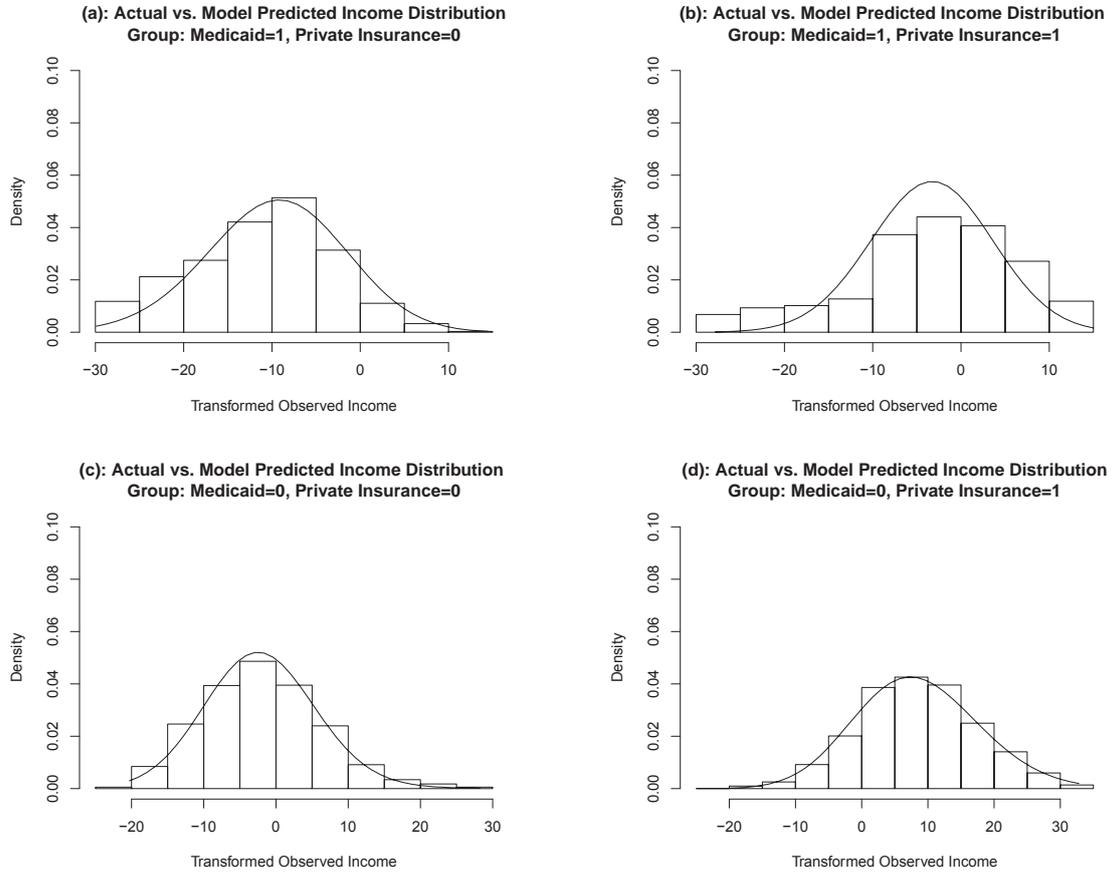

Note: Model predictions are derived using parameter estimates in Table A.3.

Table A.1: Mapping between Box-Cox Transformed and Actual Income Measures for a Child in a Four-Person Families Facing Various Medicaid Eligibility Cutoffs in 1991

|  | Medicaid Eligbility Cutoff | | | |
|---|---|---|---|---|
|  | 100% FPL | 133% FPL | 185% FPL | 200% FPL |
| Transformed Income Value | Dollar Amounts (in 1991 dollars) | | | |
| -30 | $0.01 | $1.9 | $16 | $24 |
| -15 | $147 | $250 | $442 | $503 |
| -5 | $652 | $913 | $1,341 | $1,467 |
| -1 | $1,010 | $1,356 | $1,904 | $2,063 |
| 0 | $1,117 | $1,486 | $2,066 | $2,233 |
| 1 | $1,231 | $1,623 | $2,237 | $2,413 |
| 5 | $1,764 | $2,257 | $3,015 | $3,230 |
| 15 | $3,727 | $4,527 | $5,710 | $6,039 |
| 30 | $8,806 | $10,207 | $12,210 | $12,754 |

Note: Each dollar amount represnets the actual monthly family income of a child given the value of the transformed family income and the Medicaid eligiblity cutoff she faces (the cutoff depends on the age of the child and her state of residence). The calculations above are based on a Box-Cox Transformation Parameter of 0.33, which is used in figures A.1-A.7. A dollar in 1991 is equivalent to $1.75 in 2016, and the 1991 monthly FPL for a family of four is $1117.

Table A.2: Discontinuity Estimates for Income Distribution and Medicaid and Private Insurance Coverage: Discrete Model

| Box-Cox Transformation Parameter | Percentage Trimmed | (1) Income Distribution Discontinuity | (2) Medicaid Discontinuity | (3) Private Insurance Discontinuity | (4) Estimated Crowd-out | (5) p-value from Over-ID Test |
|---|---|---|---|---|---|---|
| 0.3 | 1% | -0.017** (0.008) | -0.145*** (0.062) | 0.031 (0.122) | 0.214 (0.288) | 0.55 |
|  | 1.5% | -0.007 (0.011) | -0.226** (0.104) | 0.053 (0.140) | 0.235 (0.276) | 0.05 |
|  | 2% | -0.004 (0.007) | -0.213*** (0.083) | 0.070 (0.124) | 0.329 (0.259) | 0.06 |
|  | 2.5% | 0.001 (0.010) | -0.202** (0.100) | 0.088 (0.105) | 0.436 (0.399) | 0.09 |
| 0.33 | 1% | -0.005 (0.016) | -0.196** (0.105) | 0.067 (0.122) | 0.342 (0.360) | 0.48 |
|  | 1.5% | -0.007 (0.025) | -0.230*** (0.062) | 0.073 (0.085) | 0.317 (0.268) | 0.25 |
|  | 2% | -0.017 (0.021) | -0.201*** (0.077) | -0.016 (0.124) | -0.080 (0.357) | 0.33 |
|  | 2.5% | -0.007 (0.009) | -0.255*** (0.095) | 0.105 (0.105) | 0.412 (0.421) | 0.21 |
| 0.35 | 1% | -0.018 (0.020) | -0.193** (0.110) | 0.039 (0.098) | 0.202 (0.242) | 0.58 |
|  | 1.5% | 0.009 (0.007) | -0.205*** (0.066) | -0.029 (0.077) | -0.141 (0.282) | 0.27 |
|  | 2% | -0.003 (0.010) | -0.210*** (0.069) | 0.024 (0.082) | 0.114 (0.205) | 0.36 |
|  | 2.5% | 0.009 (0.015) | -0.241*** (0.090) | 0.020 (0.102) | 0.083 (0.187) | 0.38 |

Note: Reported are the discontinuity estimates in the true income distribution, Medicaid coverage, and private insurance coverage, respectively. Standard errors are in parentheses. *, ** and *** represent p<0.01, p< 0.05 and p<0.1 from one-sided tests with the alternative hypotheses: negative discontinuity in income distribution, negative discontinuity in Medicaid coverage and positive discontinuity in crowd-out.

Table A.3: Discontinuity Estimates for Income Distribution and Medicaid and Private Insurance Coverage: Continuous Model

|  | Point Estimate | Standard Error |
|---|---|---|
| Percent Bunching | 0.000 | (0.001) |
| Medicaid Coverage Discontinuity | -0.125*** | (0.017) |
| Crowd-out Estimate | 0.012 | (0.031) |

Note: Box-Cox Transformation parameter set to be 0.33, and the trimming percentage at 1%. Discontinuities are calculated based on maximum likelihood estimates. First stage is specified as a one-sided logit with second-order polynomial terms in transformed income variable. Outcome equation is specified as a logit with second-order polynomial terms in transformed income and the Medicaid dummy.

Table A.4: Model Predicted v.s. Actual Subgroup Proportions

| Population | Model Predicted | Actual |
|---|---|---|
| Medicaid=1, Private=0 | 11.20% | 12.00% |
| Medicaid=1, Private=1 | 2.35% | 2.12% |
| Medicaid=0, Private=0 | 14.00% | 13.20% |
| Medicaid=0, Private=1 | 72.40% | 72.70% |

Note: Model predictions are derived using parameter estimates in Table A.3.